\documentclass[sigconf]{acmart}

\usepackage{hyperref}
\usepackage{pdfpages}

\def\BibTeX{{\rm B\kern-.05em{\sc i\kern-.025em b}\kern-.08emT\kern-.1667em\lower.7ex\hbox{E}\kern-.125emX}}
\newcommand{\bill}[1]{}

\usepackage{enumitem}
\usepackage{etoolbox}
\makeatletter
\patchcmd\@combinedblfloats{\box\@outputbox}{\unvbox\@outputbox}{}{%
   \errmessage{\noexpand\@combinedblfloats could not be patched}%
}%
\makeatother

\setcopyright{none} 
\renewcommand\footnotetextcopyrightpermission[1]{}

\begin{document}

%
\title{FairST: Equitable Spatial and Temporal Demand Prediction for New Mobility Systems}

%

\author{An Yan}
\email{yanan15@uw.edu}
\affiliation{%
  \institution{University of Washington}
  \city{Seattle}
  \state{WA}
  \postcode{98105}
}

\author{Bill Howe}
\email{billhowe@uw.edu}
\affiliation{%
  \institution{University of Washington}
  \city{Seattle}
  \state{WA}
  \postcode{98105}
}



\settopmatter{printacmref=false}

\setlength{\belowcaptionskip}{-10pt}


%
\renewcommand{\shortauthors}{}

%
\begin{abstract}

Emerging transportation modes, including car-sharing, bike-sharing, and ride-hailing, are transforming urban mobility but have been shown to reinforce socioeconomic inequities. Spatiotemporal demand prediction models for these new mobility regimes must therefore consider fairness as a first-class design requirement. We present FairST, a fairness-aware model for predicting demand for new mobility systems. Our approach utilizes 1D, 2D and 3D convolutions to integrate various urban features and learn the spatial-temporal dynamics of a mobility system, but we include fairness metrics as a form of regularization to make the predictions more equitable across demographic groups. We propose two novel spatiotemporal fairness metrics, a region-based fairness gap (RFG) and an individual-based fairness gap (IFG).  Both quantify equity in a spatiotemporal context, but vary by whether demographics are labeled at the region level (RFG) or whether population distribution information is available (IFG).   Experimental results on real bike share and ride share datasets demonstrate the effectiveness of the proposed model: FairST not only reduces the fairness gap by more than 80\%, but can surprisingly achieve better accuracy than state-of-the-art yet fairness-oblivious methods including LSTMs, ConvLSTMs, and 3D CNN.


\end{abstract}

%
%



%

%

%
\maketitle

\section{Introduction}
New mobility services such as car-sharing, bike-sharing, and ride-hailing have been deployed in many cities as affordable and on-demand transportation options for citizens. For example, dockless bike share systems have been introduced in several major cities in China and the United States. They are docking-station-free and GPS-tracked. Using a mobile app, users can locate and pick up a bike closest to them, and park the bike anywhere they want \cite{mooney2019freedom, li2018free, xu2018station}. Ride-hailing companies such as Uber and Lyft connect drivers to riders through mobile phone apps. Today, they are providing over 12 million trips per day worldwide \cite{brown2018ridehail, carson2018lyft}. 

Supply and demand in new mobility systems are often unbalanced due to complex and dynamic factors such as traffic conditions and weather. Accurate and high-resolution demand estimates are therefore important to guide resource optimization and maximize system utility \cite{wang2017deepsd, Li:2018:DBR:3219819.3220110}. For example, ride-hailing companies predict demand to direct drivers to high-demand areas \cite{forecastingatUber}. Similarly, bikeshare operators use trucks to \emph{rebalance} bikes from low-demand to high-demand areas based on demand estimation \cite{mooney2019freedom}.

\begin{figure}[t]
  \centering
  \includegraphics[width=\linewidth]{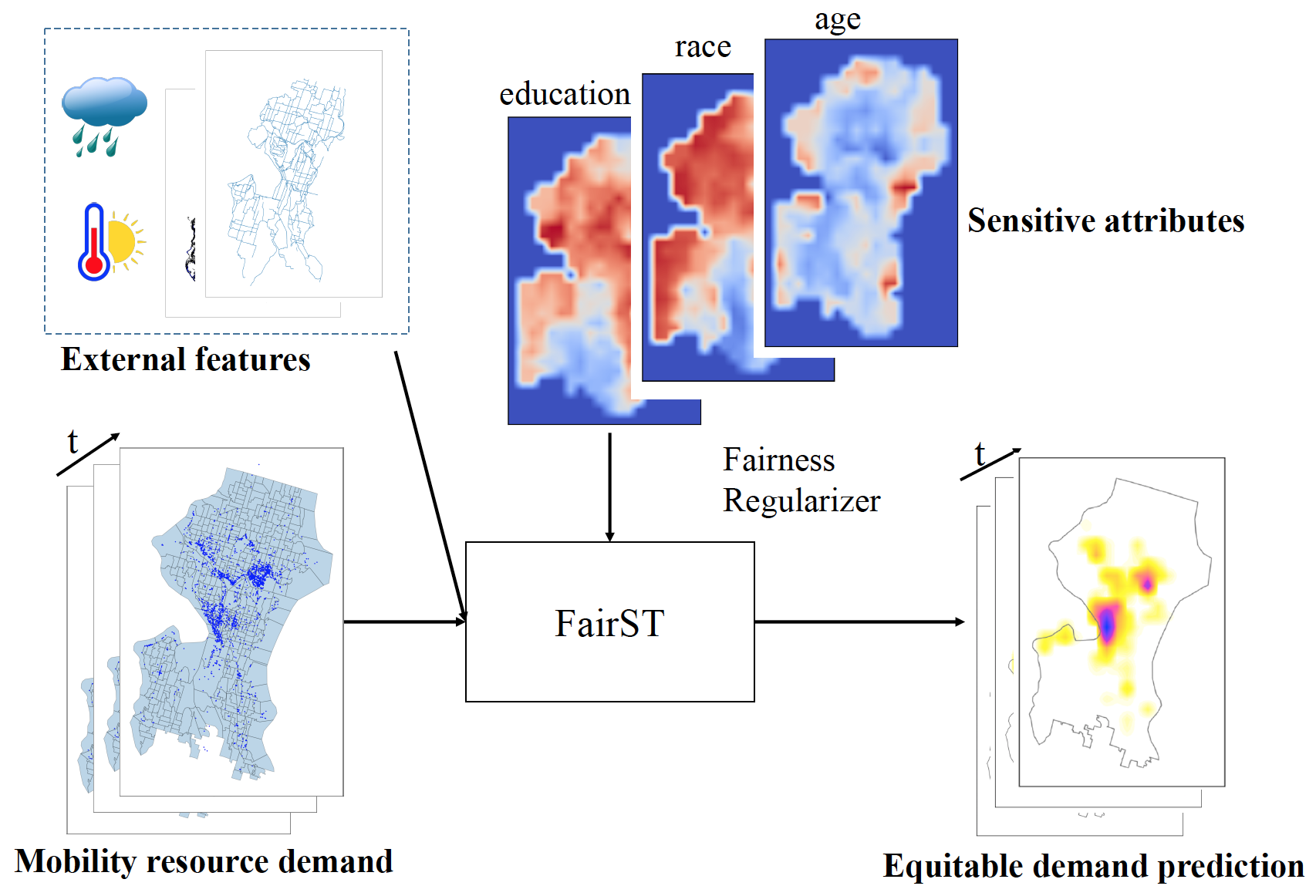}
  \caption{FairST is a deep learning based demand prediction model for new mobility systems. It not only models the spatial-temporal dynamics of mobility system, but also makes equitable predictions by incorporating a fairness regularizer that encourages equal prediction between groups defined by, for example, race, age, or education level.}
  \label{fig:teaser}
\end{figure}


Beyond accuracy, incorporating equity into demand prediction is crucial for delivering a transportation system that benefits all citizens, particularly for historically underrepresented groups. An individual's access to resources allocated or predicted by algorithms should not be dependent on sensitive attributes such as race and age. However, recent studies show that algorithms that distribute app-based mobility services may discriminate against people of color \cite{brown2018ridehail, ge2016racial}. For example, influenced by Uber's pricing algorithm, neighborhoods with more white people experienced higher service quality \cite{stark2016uber}. Compared to traditional transportation modes, new mobility services may lead to greater inequity. For example, people without smart phones are not able to access the services.  Underestimation of mobility resource demand for these groups may result in insufficient supply to these areas, which can produce a feedback loop: racial and income disparities are misinterpreted in the model as lack of demand, reinforcing reduced access to services. 




In this paper, we incorporate fairness in a demand prediction framework for new mobility systems, while acknowledging broader applications to any fair allocation of resources in space and time. To achieve our goal, the proposed approach addresses three challenges: accurate modelling of the spatial-temporal dynamics of the new mobility system, defining novel fairness metrics suitable for this task, and effective integration of fairness into the prediction model.

\textbf{Modelling the spatial-temporal dynamics of mobility resource demand.} Resource demand exhibits complex spatial and temporal patterns, and is influenced by many external factors such as weather and road network  \cite{Li:2018:DBR:3219819.3220110}. In systems such as ride-hailing and dockless bikeshare, the demand is continuous over space. Thus an initial challenge is to properly discretize continuous demand and model the spatial  dependencies among neighbouring regions. 

We address this challenge by partitioning the city into a regular grid and aggregating demand into time intervals. We use a 3D convolutional neural network (3D CNN) as the core building block in our model to capture spatial-temporal dynamics\bill{which has been shown to outperform other approaches? cite.--- No such evidence in the literature so far}. To incorporate the exogenous features such as weather and traffic that can influence demand, we adopt a three-stream model architecture that fuses together 1D, 2D, and 3D convolutional layers, respectively. A 1D CNN is used to extract information from 1D temporal features such as city-wide temperature or rainfall, and a 2D CNN is used to extract information from 2D spatial features such as the location of bike lanes.

\textbf{Designing fairness metrics for mobility resource demand}. Although the specification and assessment of fairness metrics is an active research area \cite{Glymour:2019:MBM:3287560.3287573,Hutchinson:2019:YTF:3287560.3287600}, most of the proposals are inapplicable in spatial-temporal settings. First, the prediction target in our setting (e.g., ride-hailing demand) is continuous whereas many fairness metrics such as statistical parity and equalized odds are designed for discrete classification settings \cite{Hardt:2016:EOS:3157382.3157469}. 
Second, in mobility systems, each record is usually a geographic area representing an entire subpopulation, and therefore cannot necessarily be assigned a specific attribute value (e.g. white), but rather a percentage of the subpopulation that has that value (e.g. percentage white). Among the few studies that propose fairness metrics for regression, a categorical sensitive attribute is typically required \cite{calders2013controlling,DBLP:journals/corr/BerkHJJKMNR17}. These methods cannot be directly applied to our problem unless we discretize our sensitive attributes. Finally, a fairness metric for mobility resource demand prediction should consider the overall population distribution. The transportation literature suggests that mobility resource demand is positively associated with zonal population \cite{el2017effects}, so fairness should be examined on a per capita basis. 

To address these challenges, we interpret fairness in demand prediction as the requirement that individuals of different groups have access to a similar amount of the resource in demand. We propose two fairness metrics: \textit{region-based fairness gap (RFG)} and \textit{individual-based fairness gap (IFG)}. Both assess the gap between mean per capita demand across groups over a period of time. However, RFG assumes that a distinct label is assigned to the entire region. For instance, a neighborhood with a majority white population may be assigned the label "white." IFG instead is assigned a distribution based on demographics rather than a single label.


\textbf{Integrating fairness into the prediction model.}
Fairness can be incorporated into a prediction model during data preprocessing \cite{kamiran2009classifying}, model training \cite{Dwork:2012:FTA:2090236.2090255,zafar2015fairness}, or postprocessing \cite{Hardt:2016:EOS:3157382.3157469}. During model training, fairness can be either encoded as a hard constraint or as additional terms in the loss function \cite{DBLP:journals/corr/BerkHJJKMNR17}. 
We propose two possible terms, corresponding to RFG and IFG. To the best of our knowledge, our work is the first to incorporate fairness in a spatial-temporal urban mobility setting using deep neural networks.   \bill{Are there other sptio temporal fairness papers?  If not, we can make a stronger statement here. --- I have not found any of them. }





To this end, we introduce FairST, a \textbf{Fair}ness-aware \textbf{S}patial-\textbf{T}emporal model that accounts for dynamics of mobility resource demand and enforces fairness through regularizers (Figure \ref{fig:teaser}). FairST can be naturally extended to other scenarios that involve spatial-temporal modelling and have fairness concerns such as crime incidence prediction. We summarize our main contributions as follows:
\begin{itemize}[noitemsep,topsep=0pt]
  \item We propose a new mobility resource demand prediction algorithm based on 3D convolution neural network (3D CNN) to model the temporal and spatial dependencies. The proposed algorithm adopts a three-stream architecture to integrate exogenous features with various dimensions.
  \item We propose two fairness metrics: \textit{region-based fairness gap (RFG)} and \textit{individual-based fairness gap (IFG)} for urban mobility. Both metrics measure the gap between mean per capita demand across groups over a certain period of time. The difference lies in that RFG focuses on discrete sensitive attributes while IFG deals with continuous attributes.
  \item We design and implement two fairness regularizers for deep networks in spatial-temporal settings, region-based fairness and individual-based fairness based on RFG and IFG. They are integrated into the loss minimization pipelines to encourage fair prediction. 
  \item We evaluate our method using two real-world datasets. Our experiments demonstrate that our method effectively closes the fairness gaps while achieving better accuracy than state-of-the-art fairness-oblivious models. 
\end{itemize}



\section{Related Work}

\textbf{Equity in New Mobility Systems. }A number of researchers have studied equity in bike sharing systems. Ursaki and Aultman-Hall \cite{ursaki2016quantifying} found that there are significant differences in race, education level, and income of population inside and outside bikeshare service areas in four U.S. cities. Other studies also indicate that in North America, advantaged groups have more access to docked bikeshare than disadvantaged groups  \cite{hosford2018public}. In examining access equity of dockless bikes in Seattle, Mooney et al.\cite{mooney2019freedom} found that more college-educated and higher-income residents have access to more bikes, and that bike demand is high correlated with rebalancing destinations. Overall, current literature suggests that disparities exist in the access of bikeshare systems. The equity of ride-hailing services is less clear. Although some studies found that service quality in terms of waiting times is not necessarily associated with the income or minority fraction of pickup locations \cite{hughes2016transportation, wang2018spatial}, the findings from some other studies suggest that ride-hailing companies provide poor services to low-income neighborhoods \cite{stark2016uber}. Moreover, several studies \cite{ge2016racial, brown2018ridehail} found that ride-hailing drivers discriminate against African American riders, resulting in longer waiting times and higher trip cancellation rates. 

Existing works focus mostly on assessing equity based on the outcomes of deployed systems, we argue that approaches for preventing unequal resource distribution or dynamically correcting unfairness are lacking. 




\textbf{Spatial-temporal Prediction. }
Accurate demand prediction is an essential step towards effective resource allocation (e.g., bike rebalancing and ride dispatch) strategies. Early work adopted time series analysis methods such as ARIMA or classical machine learning algorithms such as Gradient Boosting Regression Trees (GBRT) to predict mobility resource demand \cite{vogel2011understanding, yoon2012cityride, Li:2015:TPB:2820783.2820837}. Recently, deep neural networks have become popular for modeling spatial-temporal data due to their performance modeling complex non-linear interactions \cite{ZHANG2018147, wang2017deepsd}. Recurrent Neural Networks (RNN) can capture temporal dependencies \cite{gers1999learning, xu2018station} and  Convolutional Neural Networks (CNN) can capture spatial structures~\cite{Yao2018DeepMS}. Therefore, researchers use variants of RNNs and CNNs to model spatial-temporal problems \cite{liu2016predicting} such as forecasting city crowd flows \cite{ZHANG2018147}. Combinations of CNNs and RNNs were proposed to learn both temporal and spatial dependencies in one network \cite{yuan2018hetero}. ConvLSTM adopts a LSTM network structure, but replaced fully connected nodes with convolutional structures in input-to-state and state-to-state transitions, therefore achieving the advantages of CNNs and RNNs \cite{xingjian2015convolutional}. 3D Convolutional Networks were initially used for modeling video data \cite{Tran:2015:LSF:2919332.2919929}, but recently were also used for transportation demand prediction. StepDeep is a network based on 3D convolutions to predict the number of taxi trips leaving and entering a certain region of a city at a certain time. StepDeep achieved better accuracy than other methods including DeepSD \cite{shen2018stepdeep,wang2017deepsd}. 




No existing work in modeling urban resource demand considers fairness or equity in their solutions. FairST builds on the state of the art 3D CNN approaches and incorporates fair regularizers to guide the model to learn equitable spatial-temporal prediction.


\textbf{Fairness in Machine Learning.} Studies on fairness in machine learning focus on identifying and removing bias in the outcome variable with respect to some sensitive group (e.g., race, gender, income) \cite{Hutchinson:2019:YTF:3287560.3287600}. Although many competing definitions of fairness have been proposed, most involve the idea that the predicted outcomes should be statistically independent from a given sensitive attribute \cite{Dwork:2012:FTA:2090236.2090255}. Simply removing sensitive attributes from training is insufficient since there may be other features that are correlated with the sensitive attributes \cite{Zemel:2013:LFR:3042817.3042973}.   

Several fairness metrics have been proposed for classification settings. Individual fairness, in contrast to group fairness, captures the idea that similar individuals should be treated similarly \cite{Dwork:2012:FTA:2090236.2090255}. Group fairness is better aligned with most legal and practical definitions, arguing that members of a disadvantaged group should receive similar treatment to an advantaged group, by experiencing similar predicted outcomes \cite{Feldman:2015:CRD:2783258.2783311}. Equalized odds requires equal mis-classification rates across groups \cite{Hardt:2016:EOS:3157382.3157469}. Based on these concepts, researchers have proposed fairness-aware remedies that occur at all stages of the machine learning pipeline \cite{DBLP:journals/corr/BerkHJJKMNR17}. 

The majority of fairness research focuses on classification settings rather than regression settings~\cite{pmlr-v80-komiyama18a}. Metrics for classification involve discrete probabilities and are difficult to adapt directly to regression settings.  Calders et al. proposed using equal means as a fairness metric in linear regression. Fairness was incorporated through constraints in loss functions \cite{calders2013controlling}. Berk et al. developed a series of convex fairness regularizers for linear and logistic regression. They used group fairness and individual fairness analogs in regression settings. Results on six datasets highlight the incompatibility of various fairness metrics and trade-off between accuracy and fairness \cite{DBLP:journals/corr/BerkHJJKMNR17}. Our proposed method was inspired by Berk et al.'s work, but the metrics and the formulation of the loss function are novel, as is the spatial-temporal setting. 




\section{Use Cases}
In this section we describe the datasets, pre-processing, and problem formulation for our two mobility use cases.

\subsection{Datasets}

\textbf{Seattle dockless bikeshare dataset.} The city of Seattle requires shared bike operators to submit their data to the Transportation Data Collaborative (TDC) operated by the University of Washington for conducting data ethics related research. The data used in this paper comes from one of the operators from October 1, 2017 to October 31, 2018, obtained from the TDC. It includes more than 1,600,000 trips and more than 10,000 bikes. The data contains information about each bike including pickup and drop-off locations, trip start, trip end, and timestamps, as well as information about trips, including trip duration, trip start and end time, trip start location, and trip end location. We mainly use bike pick-up locations and timestamps. We use the number of pickup (trip start) as a proxy for demand as there is no ground truth value for "true demand."

\textbf{RideAustin dataset.} RideAustin \footnote{\url{http://www.rideaustin.com/}} is a non-profit ride-hailing service operating in Austin, Texas. Rides data is openly available online \footnote{\url{https://data.world/ride-austin/ride-austin-june-6-april-13}}. The data used in this paper spans from August 1, 2016 to April 13, 2017, including over 1,400,000 completed trips. It contains information about each ride including trip duration, trip start time, trip end time, trip start location, and trip end location, and distance travelled, etc. We use the number of rides as a proxy for demand.



\textbf{Socioeconomic data.} Socioeconomic data including population, race, age (under or over 65), and education level for Seattle and Austin at the block group level were obtained from the SimplyAnalytics database \cite{EASI2018}. 

\textbf{Weather features. } Previous studies show that weather conditions  are associated with bike demand and ride requests, and can be helpful for prediction \cite{Li:2015:TPB:2820783.2820837,shen2018stepdeep,wang2017deepsd}. We obtained hourly weather data for Seattle and Austin from the Integrated Surface Dataset from the National Centers for Environmental Information (NCEI) \footnote{\url{https://www.ncei.noaa.gov/access/search/index}}. We included city-level air temperature, sea level pressure, and precipitation as features for prediction. They are all 1D time series as they do not have spatial variations. 

\textbf{Urban features. } Urban forms are associated with the access and usage of new mobility systems \cite{wang2018spatial}. We collected 2D features such as bike lanes and steep slopes for Seattle bikeshare demand prediction as they may be associated with bikeshare demand according to existing literature \cite{mcneil2017breaking, frade2014bicycle, li2018free}. Likewise, we collected features such as road network and Point of Interest that were suggested by the literature for RideAustin demand prediction \cite{wang2017deepsd,shen2018stepdeep}. These urban datasets are all openly available \footnote{\url{https://data.seattle.gov/} and \url{https://data.austintexas.gov/}}. 

\subsection{Data Preparation}
\begin{figure}[h]
  \centering
  \includegraphics[width=0.7\linewidth]{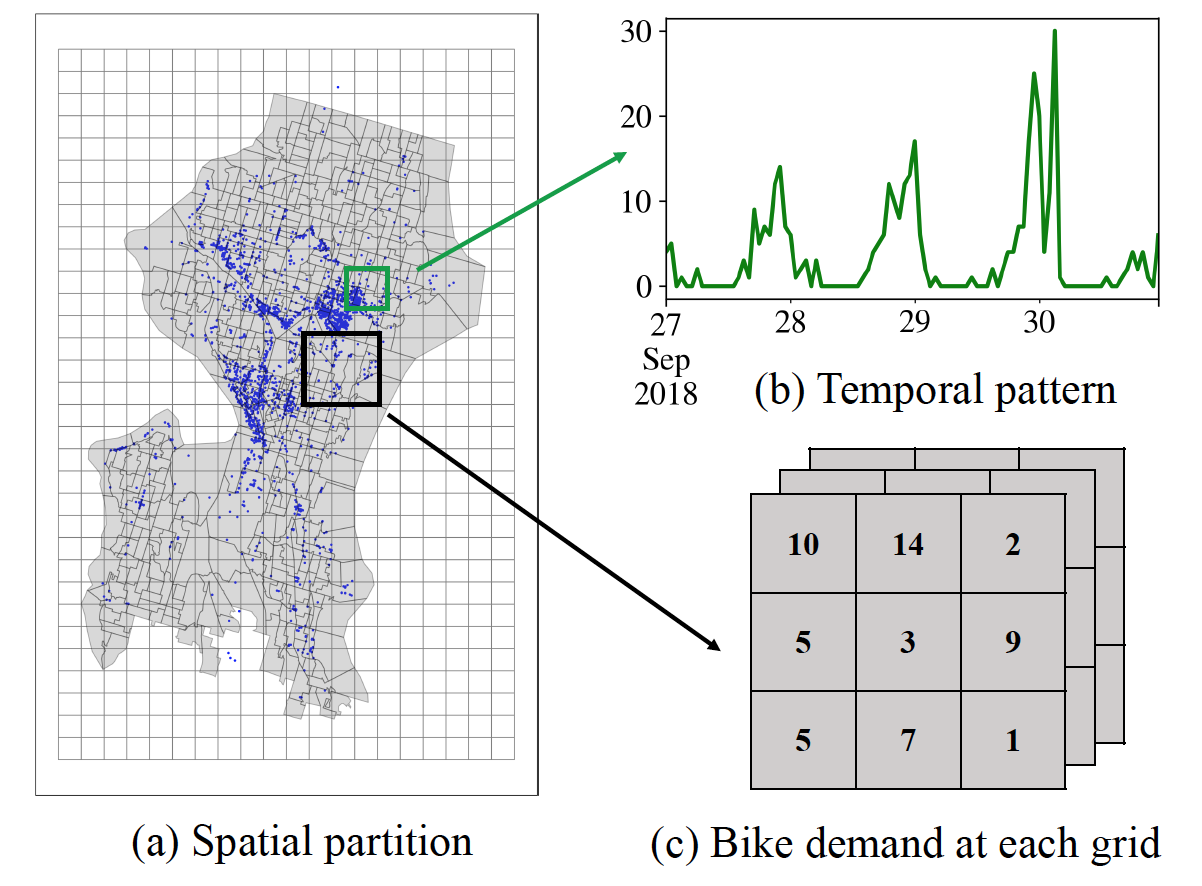}
  \caption{Data preprocessing. (a) We partition a city into square grids. (b) For each grid, resource demand forms a time series. (c) Each hour is akin to a frame in a video, with each grid cell as a pixel whose value is the demand.}
  \label{fig:data_process}
\end{figure}

Figure \ref{fig:data_process} illustrates the method that we used to process the Seattle bikeshare dataset. The RideAustin dataset was processed in the same way. We place a bounding box around the geographic region of a city and partition the bounding box into equal-sized squares (Figure \ref{fig:data_process}(a)). For Seattle bikeshare, we choose a grid size of 1km by 1km. For RideAustin, we choose a grid size of 2km by 2km. The purpose of partitioning the city into square grids rather than using neighbourhoods or block groups as the prediction unit is to prepare the data as a tensor that CNN based models can take. We counted the number of pickup in each hour and in each square region based on pickup locations and timestamps. For each grid cell, resource demand forms a time series as shown in Figure \ref{fig:data_process}(b). For each hour, the study area can be likened to a frame in a video and each region can be seen as a pixel with demand as its value (Figure \ref{fig:data_process}(c)). 

We transformed 2D urban datasets to grid cell representation using the count of features (e.g. Point of Interest, road segments) and the total length of the features (e.g. road segments) within each grid. We calculated socioeconomic attributes for each grid. Mismatches between block group boundaries and grid boundaries were accounted for using proportional allocation based on area.


\subsection{Prediction Problem Definition}
We aim to build fair models to forecast next time step demand for mobility resource for a city based on the demand of previous time steps. For both Seattle bikeshare and RideAustin, we aim to predict hourly demand based on the demand of the last 7 days (168 hours). The prediction problem is similar to predicting next frame based on the previous 168 frames in a video. We generated slices of 169 hours for training and prediction (168 hours for training and to predict the next 1 hour). For Seattle bikeshare, we use the data from October 2017 to August 2018 for training and the data from September to October, 2018 for testing. The training data contains 8040 temporal slices and the test data contains 1464 temporal slices. For RideAustin, we use the data from August 2016 to February 2017 for training and the data from March to April 2017 for testing. The training data contains 5088 temporal slices and the test data contains 1056 slices. The prediction should balance two objectives: minimizing prediction accuracy loss and minimizing fairness loss. 




\section{Model and Fairness Metrics}
In this section, we detail our spatial-temporal model architecture and describe our proposed fairness metrics and corresponding fairness regularizers.  

\subsection{Model Architecture} \label{Model}
We first introduce 3D convolutions for learning spatial-temporal features, then present the architecture of FairST, followed by the design of the objective function that guides the learning process. 

\textbf{3D convolutions} The core building block of FairST is 3D convolution, which models spatial-temporal information. 2D CNNs use 2D convolution operations to compute features from spatial dimensions. When performing 2D convolutions on multiple channels of an image or multiple images, the output is a one-dimensional image. As the input of spatial-temporal slices are 3D tensor like frames in videos, 2D convolution will result in a 2D output without temporal information. 3D convolution can preserve temporal information as the multiple contiguous frames of input are connected by the 3rd dimension of the filters, therefore the output is another 3D tensor containing temporal information \cite{ji20133d}.





\begin{figure*}
  \centering
  \includegraphics[width=0.85\linewidth]{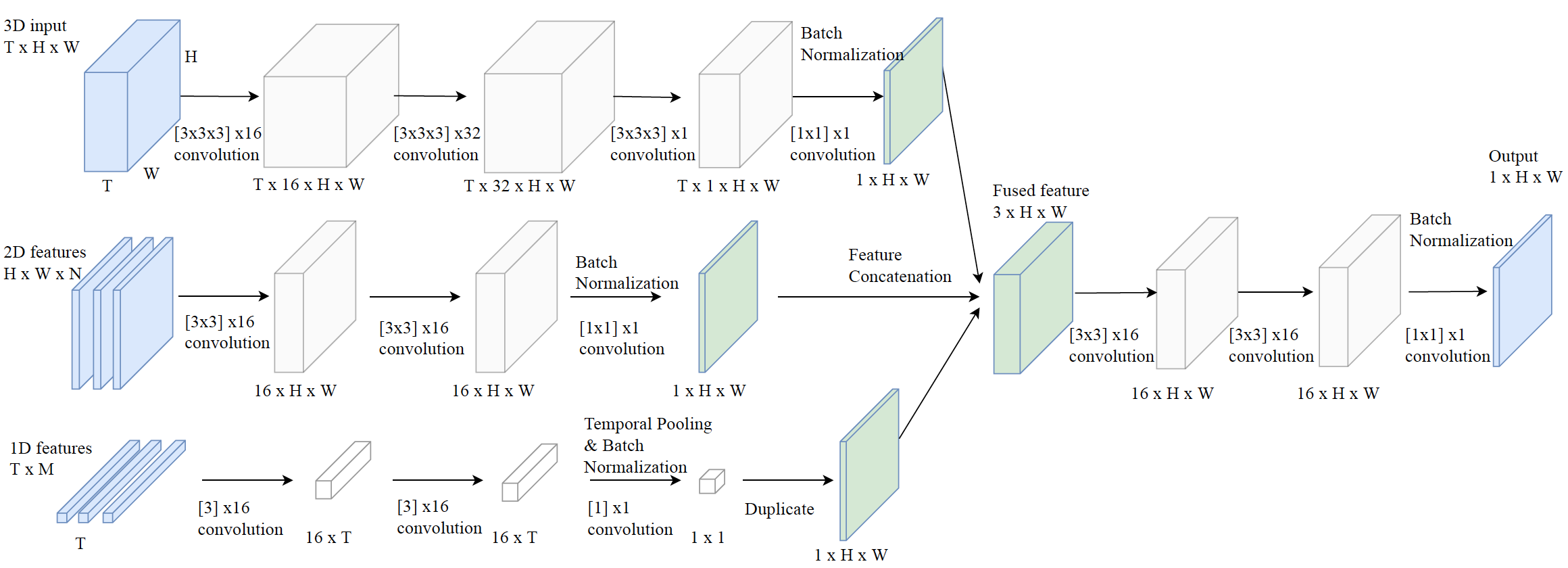}
  \caption{A three-stream network architecture. The network input contains three streams, including 1D time series features, 2D urban features, and 3D spatial-temporal input. The network is trained to predict mobility resource demand in an end-to-end fashion. T, H, W are the number of time steps, height of input, and width of input, respectively. N and M are the number of 2D and 1D features, respectively.}
  \label{fig:structure}
\end{figure*}

\textbf{Network architecture}. We design a three-stream prediction framework based on 1D, 2D, and 3D CNN to 1) automatically capture the spatio-temporal context, and 2) include external features to help with accuracy. We use a submodel consists of 3D convolution layers to learn from 3D historical demand, a submodel with 1D convolution layers to learn information from 1D time series features, and a submodel with 2D convolution layers to extract information from 2D urban features. The outputs of all submodels were fused together, on top of which additional convolutional layers were applied to achieve the final prediction (See Figure \ref{fig:structure}). Compared to fusing all features before being fed to a single network, this strategy has two main advantages: 1) Integrating semantically related features into one submodel can potentially reinforce the effectiveness of one another \cite{zheng2014urban}. For example, in our setting, 1D features often represent mutually correlated meteorological information, and 2D features reflect the geographic characteristics of the city. 2) Fusing all features early at the dataset level requires 1D and 2D features to be replicated to create 3D tensors. This redundancy brings unnecessary computation overhead and wasted model capacity.

The first submodel is based on 3D convolutions. It takes a time series of resource demand history as input. The submodel consists of three 3D convolutional layers, followed by a 2D convolutional layer, as shown in Figure \ref{fig:structure}. The number of filters of 3D convolutional layers are 16, 32, and 1, respectively. We use 3 $\times$ 3 $\times$ 3 filters because this is the size that worked best in previous studies \cite{Tran:2015:LSF:2919332.2919929}. We use padding to ensure the layer outputs are of the same size as inputs. The third 3D convolutional layer adopts 1 filter to achieve dimension reduction \cite{szegedy2015going} and temporal pooling. Finally, a 2D convolution layer is used to integrate temporal information from previous layers and output the feature map for submodel fusion. The second and third submodels are based on 2D and 1D convolutions, respectively. They aim to extract meaningful information from the input features and improve prediction accuracy\bill{I don't understand what the point is here. --- I wanted to say sth about each submodel, but wanted to say more about 3D CNN. Maybe we could remove these two sentence...}. We use leaky Relu as activation function for all layers \cite{maas2013rectifier}. We keep the model light-weight and skip spatial pooling to avoid deconvolution operations (for maintaining the output size) afterwards, which is typically more prone to overfitting in small training sets.



\textbf{Training objectives}. Our loss function is a weighted sum of an accuracy loss and a fairness loss. The fairness loss acts as a regularizer for the model. We use Mean Absolute Error (MAE) as accuracy loss. The overall loss function is defined as



\begin{equation}
  L = L_{accuracy} + \lambda L_{fairness}
\end{equation}

where $L_{accuracy}$ is MAE, $L_{fairness}$ is the fairness loss, and $\lambda$ is the weight for the fairness loss. In the experiments (Section \ref{sec:Experiments}), we show the accuracy given different $\lambda$ values. In the next section, we describe details of the proposed fairness loss. 

\subsection{Fairness Metrics and Regularizers}
We consider fairness as individuals of different groups receiving equal resources. In the mobility setting, fair prediction implies adjusting the demand to reduce the difference in per capita resource demand among groups defined by, say, race. Our definition adapts \emph{group fairness} in the machine learning literature that requires the disadvantaged group to receive similar treatment to the advantaged group by experiencing similar predicted outcomes \cite{Feldman:2015:CRD:2783258.2783311}, and is informed by \emph{vertical equity} in the transportation literature requiring transportation policies to favor socially disadvantaged groups to compensate for overall inequities \cite{delbosc2011using}.

Given this approach to fairness, we propose two fairness metrics: a Region-based Fairness Gap (RFG) and an Individual-based Fairness Gap (IFG).  Both RFG and IFG measure the gap between mean per capita demand across two groups over a certain period of time. However, for RFG, each geographic region is assigned a single group label according to some criteria (e.g., Caucasian or non-Caucasian). For IFG, groups are determined based on the demographic distribution  in the region, such that the sensitive attribute is numeric (e.g., the percentage of the subpopulation in the region that is Caucasian). In this paper we focus on a square grid partitioning, these two metrics can be used for any customized partitioning (e.g., census tracts, zip codes, etc.)

\textbf{Intuition.} RFG draws upon the idea that people live in the same region share similar public facilities and economic status, so they may have similar commute patterns and demand for transportation resources. For example, a white person may live in a predominately black community, but she frequents the same bus stops and grocery stores as her neighbors.\bill{This argument is compelling, but is there something to cite that makes a similar argument?} Therefore, when assessing mobility resource demand equity, policies to distribute resources may primarily consider the majority group. In practice, we can assign each region the group label (e.g., race) with the highest population, or some criteria defined by local governments. However, we caution that a simple discretization of the sensitive attributes by a threshold for each region itself is biased, since the minority population in a region may be underrepresented.

\textbf{Notation.} We start by introducing notation.
\begin{itemize}[noitemsep,topsep=0pt]
    \item Let $s_{i}$ be the $i$th square region of the study area $\mathcal{S}$.
    \item Let $p_{i}$ denote the population of square region $s_{i}$ divided by the total population of the city. 
    \item Let $\hat{y}_{i,t}$ and ${y}_{i,t}$ be the estimated demand and ground truth demand for region $s_{i}$ at time $t$, respectively. 
    \item Let $E_{T}[\hat{y}_{i,t}]$ be the average predicted value for the $i$th square region in $\mathcal{S}$ over time period $T$.
\end{itemize}


\textbf{Region-based Fairness Gap (RFG)}. 
We now formally define RFG. Let every region $s_{i}$ be assigned to one of two groups (e.g., Caucasian and non-Caucasian) with regard to one sensitive attribute $A$ (e.g., race), denoted by $G^+$ (the advantaged group) and $G^-$ (the disadvantaged group). We define RFG between two groups with regard to sensitive attribute $A$ over a period of time $T$ as follow: 



\begin{equation}
  RFG =  \frac{\sum_{i \in G^+} E_{T}[\hat{y}_{i,t}]}{\sum_{i \in G^+}p_{i}} -\frac{\sum_{j \in G^-} E_{T}[\hat{y}_{j,t}]}{\sum_{j \in G^-} p_{j}} 
\end{equation}

The first term can be interpreted as the per capita demand for group $G^+$ averaged over $T$. The denominator is the total population (normalized) of $G^+$. Likewise, the second term is the mean per capita demand in group $G^-$ over $T$.

\textbf{Individual-based Fairness Gap (IFG)}. Let $w_{i}^+$  denote the percentage of people in the advantaged group of the sensitive attribute $A$ (e.g., race) in region $s_{i}$ and let $w_{i}^-$ denote the percentage of people in the disadvantaged group. For example, if a region $s_{i}$ is 65\% white, then $w_{i}^+$ = 65\% and $w_{i}^-$ = 35\%. IFG assumes that given the predicted demand, the number of resources a group will get is proportional to the population percentage of that group. For example, if the predicted demand for bikeshare is 100 bikes for a region and the percentage of white people is 65\%, then the demand that allocated to the Caucasian group in that region is 65 bikes. Formally, we define IFG between two groups with regard to sensitive attribute $A$ over a period of time $T$ as follow:

\begin{equation}
  IFG =  \frac{\sum_{i \in \mathcal{S}} E_{T}[\hat{y}_{i,t}]w_{i}^+ }{\sum_{i \in \mathcal{S}}p_{i}w_{i}^+} -\frac{\sum_{j \in \mathcal{S}} E_{T}[\hat{y}_{j,t}] w_{j}^-}{\sum_{j \in \mathcal{S}} p_{j}w_{j}^-} 
\end{equation}

The numerator of the first term denotes the predicted total demand of all people in the advantaged group averaged over $T$. The denominator is the total population (normalized). Then the first term is the predicted per capita demand allocated to the advantaged group averaged over $T$. The second term can be interpreted similarly.  

In summary, for RFG, everyone that lives in the same region is assigned the same group label, whereas IFG assigns group labels proportionally based on the region's demographics. 


\textbf{Fairness loss}. Based on the RFG and IFG, we define two fairness loss terms, Region-based Fairness loss (RF loss) and Individual-based Fairness loss (IF loss) to incorporate fairness into training. 

The \emph{Region-based Fairness loss (RF loss)} at time $t$ is defined as 

\begin{equation}
  L_{RF} (t) = \frac{1}{\sum_{i \in \mathcal{S}} y_{i,t} }   \left| \frac{\sum_{i \in G^+}\hat{y}_{i,t}}{\sum_{i \in G^+}p_{i}} -\frac{\sum_{j \in G^-}\hat{y}_{j,t}}{\sum_{j \in G^-} p_{j}} \right|
\end{equation}

The first term is the estimated per capita demand in group $G^+$ at time $t$. Likewise, the second term is for group $G^-$.  $\sum_{i \in \mathcal{S}} y_{i,t} $ is a normalizing factor.

\emph{The Individual-based Fairness loss (IF loss)} at time $t$ is defined as 


\begin{equation}
  L_{IF} (t) = \frac{1}{\sum_{i \in \mathcal{S}} y_{i,t} }   \left| \frac{\sum_{i \in \mathcal{S}}\hat{y}_{i,t} w_{i}^+}{\sum_{i \in \mathcal{S}}p_{i}w_{i}^+} -\frac{\sum_{j \in \mathcal{S}}\hat{y}_{j,t}w_{j}^-}{\sum_{j \in \mathcal{S}} p_{j}w_{j}^-} \right|
\end{equation}

The first term is the estimated per capita demand for advantaged group at time $t$. Likewise, the second term is for disadvantaged group. 

\emph{Multiple sensitive attributes} can be represented together in one loss function as the weighed sum of fairness loss of each attribute. Assuming there are $a = \{1,2,..., A \}$ sensitive attributes, then the composite loss function is defined as

\begin{equation}\label{eq:multivar}
  L_{fairness} (t) = \sum_{a = 1}^{A}  \lambda_{a}  L_{fairness_{(a,t)}}
\end{equation}

where $\lambda_{a}$ is the weight term for the $a$th attribute and $L_{fairness_(a,t)}$ is the fairness loss. 


\section{Experiments} \label{sec:Experiments}
We evaluate our method on the Seattle dockless bikeshare dataset and the RideAustin dataset. First, we compare FairST without fairness loss ($\lambda = 0$) with state-of-the-art spatial-temporal models in terms of prediction accuracy. We then incorporate Region-based Fairness loss (RF loss) and Individual-based Fairness loss (IF loss) into our model. To understand the effectiveness of the two proposed fairness regularizers, we compare against other existing fairness regularizers on a single sensitive attribute (i.e. race). Finally, we integrate the fairness losses for race, age, and education level into FairST to evaluate its capability of reducing unfairness for multiple sensitive attributes in one shot. 

\subsection{Implementation} \label{sec: implementation}

We implement FairST and the deep-learning based baseline models with TensorFlow Framework \cite{abadi2016tensorflow}, and perform training and inference with NVIDIA K80 GPU machines. We use a batch size of 32 and train FairST for 200 epochs for Seattle bikeshare and 350 epochs for RideAustin using Adam optimizer. We use a exponential learning rate decay scheme: the learning rate starts at 0.005 and decays every 5,000 steps with a rate of 0.96. 

To implement Region-based Fairness loss, we assign each square region a label for each attribute. We use the overall city statistics as thresholds to discretize the continuous sensitive attributes. For example, the percentage of white population of Seattle in 2018 is 65.74\%, we then set the regions with more than 65.74\% white population as Caucasian group, otherwise as non-Caucasian group. The same method is used for discretizing age and education level. 

\subsection{Baseline Models}
To evaluate the prediction accuracy of our method, we compare FairST with several other models: 1) \textbf{Historical Average (HA)}. We compute $\hat{y}_{i, t}$ using the mean values of all previous observations at location $s_{i}$ at the same time of the day and the same day of the week. 2) \textbf{Autoregressive Integrated Moving Average Model (ARIMA)}. ARIMA is a commonly used statistic model for forecasting time series. We develop an independent ARIMA model for each individual grid cell. 3) \textbf{Long short-term memory Network (LSTM)} \cite{gers1999learning}. LSTM is a variant of Recurrent Neural Network that can learn long-term temporal dependencies. We train the LSTM model individually for each square grid. 4) \textbf{Convolutional LSTM (ConvLSTM)} \cite{xingjian2015convolutional}. The ConvLSTM network adopts LSTM structure, but replaces fully connected layers with convolutional operations in each cell. As a result, it can capture both spatial and temporal dependencies in one network. We also compare FairST with various 3D CNN models: a \textbf{3D CNN} model that is equivalent to FairST without any external features; a \textbf{3D CNN + 1D} model that consists of a 3D CNN based submodel and a 1D CNN based submodel; and a \textbf{3D CNN + 2D} model that consists of a 3D CNN based submodel and a 2D CNN based submodel.

\subsection{Baseline Fairness Regularizers}
We compare the proposed loss functions (RF loss and IF loss) with two other existing fairness losses \cite{calders2013controlling, DBLP:journals/corr/BerkHJJKMNR17} in our experiments. 


\textbf{Equal Means Loss (EM Loss)}. Calders et al. defined Equal Means as a fairness metric for regression \cite{calders2013controlling}. Equal Means enforces the mean prediction to be the same for different groups. This metric is not directly comparable with IFG or RFG as we focus on predicted demand per capita, therefore, we substitute prediction with per capita prediction in Equal Means loss. The modified Equal Means loss is defined as:


\begin{equation}
  L_{EM} (t) =  \frac{1}{\sum_{i \in \mathcal{S}} y_{i,t} }   \left| \frac{\sum_{i \in G^+} \hat{z}_{i,t}}{n^{+}}    -  \frac{ \sum_{j \in G^-}\hat{z}_{j,t}}{n^-} \right|
\end{equation}


where $p_{i}$ is the population of region $s_{i}$ divided by the total population of the city. $\hat{z}_{i,t} = \frac{\hat{y}_{i,t}}{p_{i}}$, denoting the predicted per capita demand. $n^{+}$ and $n^-$ denote the number of advantaged square regions and the number of disadvantaged square regions, respectively.

\textbf{Pairwise Fairness Loss (Pairwise Loss)}. Berk et al. defined a family of fairness regularizers that corresponds to individual fairness, group fairness, and hybrid of the two \cite{DBLP:journals/corr/BerkHJJKMNR17}. In all three loss term formations, comparisons across groups are based on cross pairs $i \in G^+$ and ${j \in G^-}$. Since our metrics are analogs of group fairness, we compare our metrics with Berk's group fairness penalty. 

\begin{equation}
  L_{PF} (t) = \frac{1}{\sum_{i \in \mathcal{S}} y_{i,t} }   
 \bigg( \frac{1}{n^+ n^-} \sum_{\substack{{i \in G^+} \\ {j \in G^-}}}
  d({z}_{i,t}, {z}_{j,t})  (\hat{z}_{i,t} -\hat{z}_{j,t}) \bigg) ^2
\end{equation}

\begin{equation}
 d({z}_{i,t}, {z}_{j,t}) = e^{- ({z}_{i,t} -{z}_{j,t})^2}
\end{equation}

Similar to the modified Equal Means loss, we substitute prediction with per capita prediction. The model will increase penalty as the difference between $\hat{z}_{i,t}$ and $\hat{z}_{j,t}$ increases, weighted by a similarity function $ d({z}_{i,t}, {z}_{j,t}) $. 


\subsection{Evaluation Metrics}
We evaluate the prediction accuracy of all models with Mean Absolute Error (\textbf{MAE}). We evaluate the fairness of prediction outcomes using \textbf{RFG} and \textbf{IFG}, but we also consider the correlation between the ranked demand and the proportion of the disadvantaged group. That is, we are considering that city planners are interested in assessing whether the regions with the highest demand also happen to be the wealthy, advantaged neighborhoods. We use Spearman's rank correlation coefficient (\textbf{Spearman's rho}) \cite{hauke2011comparison}, which measures the strength of monotonic correlation between two variables.  We calculate Spearman's rho between mean per capita demand over the test period of a grid region and the percent of advantaged population (i.e., percentage of Caucasian, percentage of population under 65 years old, and percentage of population with a college degree) of that region. A highly positive Spearman's rho with a p-value less than 0.05 suggests disparities in demand.





\section{Results and Discussion}
The primary goal of predicting demand is to guide resource allocation, so it is desirable to make accurate predictions while closing the equity gaps. In this section, we show that proposed fairness regularizers give better performance than baseline regularizers in our problem setting. We also show that FairST is able to achieve better accuracy and less inequity than baseline models. 

\subsection{Demand Prediction Accuracy}
We compare prediction accuracy of our model with baselines. Table \ref{tab:mae_fairness} and Table \ref{tab:sing_var_fairness_austin} show Mean Absolute Error of all models on the Seattle bikeshare dataset and the RideAustin dataset, respectively. It is observed that the 3D CNN based methods (i.e., 3D CNN, 3D CNN + 1D, 3D CNN + 2D, and FairST without fairness penalty) proposed by this paper achieve higher prediction accuracy than the other methods. HA is a simple and reasonable method to predict demand, however, it overgeneralizes temporal dynamics and does not account for spatial structure. ARIMA assumes input time series is stationary which is often not the case with fluctuating demand. It is also not good at predicting with sparse data where there are many zeros in the series. LSTM achieves better accuracy than ARIMA and HA, but still suffers from inability to learn information from spatial context. ConvLSTM outperforms LSTM due to its capability of learning both spatial and temporal information. The 3D CNN models perform better than ConvLSTM since the 3D CNN is more powerful in terms of capturing strong local spatial-temporal correlations in our problem as compared to the recurrent architectures. Furthermore, the incorporation of external features improves accuracy in both Seattle bikeshare and RideAustin cases.

\begin{table*}

\caption{FairST compared to baselines for predicting Seattle bikeshare demand (multiple attributes)}
\label{tab:mae_fairness}
\addtolength{\tabcolsep}{-1.9pt}  
\begin{tabular}{lllllllllllllll}
\toprule
             & $\lambda$ & MAE   & \begin{tabular}[c]{@{}l@{}}RFG \\ (race)\end{tabular} & \begin{tabular}[c]{@{}l@{}}RFG\\ (age)\end{tabular} & \begin{tabular}[c]{@{}l@{}}RFG\\ (edu)\end{tabular} & \begin{tabular}[c]{@{}l@{}}IFG\\ (race)\end{tabular} & \begin{tabular}[c]{@{}l@{}}IFG\\ (age)\end{tabular} & \begin{tabular}[c]{@{}l@{}}IFG\\ (edu)\end{tabular} & \begin{tabular}[c]{@{}l@{}}Spearman's  \\ rho (race)\end{tabular} & \begin{tabular}[c]{@{}l@{}}Spearman's  \\ rho (age)\end{tabular} & \begin{tabular}[c]{@{}l@{}}Spearman's  \\
              rho (edu)\end{tabular} \\
\midrule
Ground Truth & /      & /     & 112.568                                               & 160.089                                             & 37.471                                              & 38.969                                               & 51.338                                              & 30.053                                              & 0.016                                                      & 0.174$^{**}$                                                   & 0.338$^{**}$                                                                                             \\
HA           & /      & 0.484 & 194.454                                               & 49.494                                              & 193.477                                             & 79.906                                               & 17.641                                              & 54.692                                              & 0.565$^{**}$                                                    & 0.477$^{**}$                                                   & 0.500$^{**}$                                                                                               \\
ARIMA        & /      & 0.538 & 319.032                                               & 62.793                                              & 319.648                                             & 129.447                                              & 28.170                                              & 90.505                                              & 0.569$^{**}$                                                    & 0.463$^{**}$                                                   & 0.489$^{**}$                                                                                                  \\
LSTM\cite{gers1999learning}         & /  & 0.468 & 280.685                                               & 61.437                                              & 277.938                                             & 116.023                                              & 23.778                                              & 79.162                                              & 0.522$^{**}$                                                    & 0.441$^{**}$                                                   & 0.425$^{**}$                                                                                \\
ConvLSTM \cite{xingjian2015convolutional}    & 0.000  & 0.432 & 74.485                                                & 139.666                                             & 19.934                                              & 22.907                                               & 44.459                                              & 19.101                                              & 0.210$^{**}$                                                & 0.355$^{**}$                                                  & 0.324$^{**}$                                                                                    \\
3D CNN       & 0.000  & 0.408 & 100.878                                               & 169.240                                             & 38.873                                              & 31.915                                               & 53.133                                              & 26.851                                              & 0.091                                                      & 0.256$^{**}$                                                  & 0.394$^{**}$                                                                                                 \\
3D CNN + 1D  & 0.000  & 0.387 & 88.587                                                & 153.625                                             & 19.802                                              & 26.791                                               & 49.058                                              & 20.691                                              & 0.291$^{**}$                                                    & 0.376$^{**}$                                                   & 0.077                                                                                               \\
3D CNN + 2D  & 0.000  & 0.378 & 93.299                                                & 157.025                                             & 33.946                                              & 28.661                                               & 49.792                                              & 24.457                                              & 0.158$^{**}$                                                    & 0.246$^{**}$                                                   & 0.370$^{**}$                                                                                           \\
FairST       & 0.000  & 0.382 & 83.127                                                & 147.437                                             & 23.400                                              & 25.073                                               & 47.403                                              & 20.885                                              & 0.168$^{**}$                                                    & 0.191$^{**}$                                                   & 0.328$^{**}$                                                                                          \\
FairST + RF  & 0.005  & \textbf{0.377} & 80.565                                                & 146.665                                             & 20.855                                              & 24.168                                               & 46.732                                              & 20.184                                              & 0.111$^{*}$                                                      & 0.262$^{**}$                                                   & 0.348$^{**}$                                                                                                 \\
FairST + RF  & 0.150  & 0.437 & 16.140                                                & 35.562                                              & -5.712                                              & 4.199                                                & 22.543                                              & 7.112                                               & -0.019                                                     & 0.107$^{*}$                                                    & 0.321$^{**}$                                                                                                 \\
FairST + RF  & 0.250  & 0.460 & \textbf{8.650}                                                & \textbf{14.242}                                              & \textbf{-3.364}                                              & 2.226                                                & 19.178                                              & 6.299                                               & \textbf{0.011}                                                      & \textbf{0.090}                                                     & 0.231$^{**}$                                                                   \\
FairST + IF  & 0.100  & 0.385 & 67.695                                                & 128.010                                             & 4.905                                               & 17.927                                               & 40.811                                              & 14.874                                              & 0.099                                                      & 0.231$^{**}$                                                   & \textbf{0.347$^{**}$}                                                                                               \\
FairST + IF  & 0.150 &  0.394  & 49.075  &	110.725  &	-9.322  &	11.738 &	35.410& 	9.529 &  0.030  &  0.181$^{**}$  & 0.385$^{**}$ \\ 
FairST + IF  & 0.500  & 0.439 & 30.668                                                & 53.896                                              & -20.291                                             & 3.823                                                & 16.536                                              & 2.200                                               & 0.117$^{*}$                                                      & 0.222$^{**}$                                                   & 0.085                                                                                                   \\
FairST + IF  & 0.600  & 0.460 & 24.753                                                & 34.011                                              & -22.700                                             & \textbf{0.898}                                                & \textbf{8.855}                                               & \textbf{-0.185}                                              & 0.060                                                      & 0.158$^{**}$                                                   & \textbf{-0.055}                                                                                                    \\ 
\bottomrule
\multicolumn{12}{l}{\footnotesize{$^{**}$. Correlation is significant at the 0.05 level. }}\\
\multicolumn{12}{l}{\footnotesize{$^*$. Correlation is significant at the 0.01 level.}}\\
\end{tabular}
\addtolength{\tabcolsep}{1.9pt}
\end{table*}





\subsection{Fair Prediction: Single Attribute}
\begin{table}[h]
  \caption{FairST compared to baselines for Seattle bikeshare demand prediction (single attribute)}
  \label{tab:sing_var_fairness_seattle}
\addtolength{\tabcolsep}{-2pt} 
\begin{tabular}{lllllll}
\toprule
            & $\lambda$ & MAE   & \begin{tabular}[c]{@{}l@{}}RFG \\ (race)\end{tabular} & \begin{tabular}[c]{@{}l@{}}IFG\\ (race)\end{tabular} & \begin{tabular}[c]{@{}l@{}}Spearman's\\rho (race)\end{tabular} & \\
\midrule
ConvLSTM\cite{xingjian2015convolutional}    & 0.00  & 0.432 & 74.485                                                & 22.907                                               & 0.210$^{**}$                                                                                                     \\
3D CNN      & 0.00  & 0.408 & 100.878                                               & 31.915                                               & 0.091                                                                                                        \\
FairST      & 0.00  & 0.382 & 83.127                                                & 25.073                                               & 0.168$^{**}$                                                                                                   \\
FairST + RF & 0.02  & \textbf{0.379} & 79.570                                                & 24.694                                               & 0.144$^{**}$                                                                                                \\
FairST + RF & 0.50  & 0.404 & 10.627                                                & 3.363                                                & -0.076                                                                                                   \\
FairST + RF & 0.90  & 0.440 & 0.017                                                 & 0.473                                                & \textbf{-0.005}                                                                                                    \\
FairST + IF &0.20 & \textbf{0.379}  &	63.130&	15.281	&0.085 \\

FairST + IF & 1.50  & 0.406 & 38.473                                                & 4.902                                                & -0.070                                                                                                   \\
FairST + IF & 5.00  & 0.442 & \textbf{0.004}                                                 & \textbf{0.266}                                                & -0.046                                                  \\                               \bottomrule  
\multicolumn{5}{l}{\footnotesize{$^{**}$. Correlation is significant at the 0.05 level. }}\\
\multicolumn{5}{l}{\footnotesize{$^*$. Correlation is significant at the 0.01 level.}}\\
\end{tabular}
  \addtolength{\tabcolsep}{2pt} 
\end{table}

\begin{figure}
  \centering
  \includegraphics[width=\linewidth]{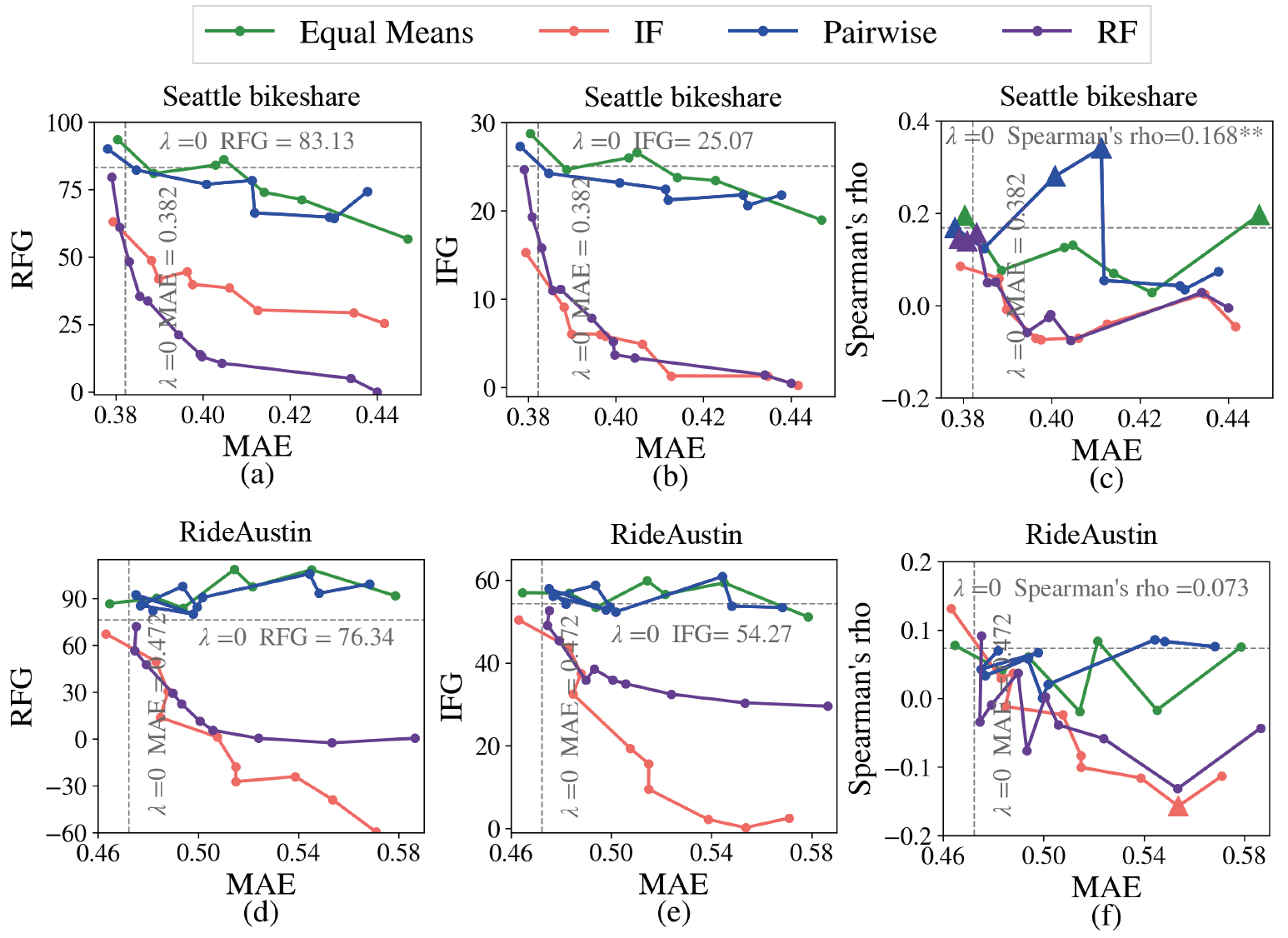}
  \caption{Accuracy vs. fairness metrics (single attibute). (a), (b), and (c) show the relationship between MAE vs. RFG, IFG, and Spearman's rho,  respectively for Seattle bikeshare. (d), (e), and (f) show the results of RideAustin. Triangles in (c) and (f) represent statistical significance (p-value < 0.01).}\label{fig:sing_var}
\end{figure}

We trained FairST with Region-based Fairness loss (RF), Individual-based Fairness loss (IF), Equal Means loss (Equal Means), and Pairwise loss (Pairwise) respectively, on a single attribute, i.e. race on two datasets. Figure \ref{fig:sing_var} illustrates the relationships between MAE and  fairness metrics, each point on a curve corresponds to a $\lambda$ value, which increases from left to right of the curve.

Figure \ref{fig:sing_var} (a), (b), (d), and (e) show that RF and IF regularizer are very effective in controlling both RF and IF gaps. Overall, we observe a trade-off between MAE and IFG (or RFG). That is, accuracy decrease as fairness regularizer strength ($\lambda$) increases. In Seattle bikeshare case, IF regularizer ($\lambda$ = 0.2) brings IFG down from 25.073 to 15.281 while keeping better accuracy than FairST with $\lambda$ = 0 (see Table \ref{tab:sing_var_fairness_seattle}). This also suggests that the use of fairness loss terms ($\lambda > 0$) may \emph{improve} the MAE over the baseline model ($\lambda = 0$) for small values of $\lambda$. The reason is that the addition of fairness terms provides a regularizing effect on accuracy. In contrast, with the Equal Means regularizer or the pairwise regularizer, the models show no clear patterns in terms of RFG or IFG. 


Figure \ref{fig:sing_var} (c) and (f) show the fairness of models evaluated by Spearman's rho. Triangles represent statistical significance (p-value < 0.01). In Seattle bikeshare case, FairST ($\lambda = 0$) without fairness would result in an unfair prediction (see Table \ref{tab:sing_var_fairness_seattle}). That is, there is a positive monotonic correlation (Spearman's rho = 0.168, p-value < 0.01) between the predicted demand and the percent of Caucasian population. Models with an IF or a RF regularizer effectively bring down the Spearman's rho to around zero, and the predictions are no longer significantly correlated with race as $\lambda$ increases. In contrast, Spearman's coefficients of models with an Equal Means regularizer or a pairwise regularizer stay positive throughout and sometimes show significantly positive correlation between the predicted outcome and race. In the RideAustin case, the predicted outcome of FairST ($\lambda = 0$) does not show a significant correlation with race. The Spearman's coefficients of models with an IF regularizer or a RF regularizer decrease and stay below zero, while the patterns of models with an Equal Means regularizer or a pairwise regularizer are less clear. 

    

Table \ref{tab:sing_var_fairness_seattle} shows the results of FairST compared to baselines for Seattle bikeshare. Both the RF regularizer and IF regularizer bring down about 85\% IFG (from 31.915 to 3.363 and 4.902, respectively) while keeping better MAE than 3D CNN (MAE = 0.408). They also bring down IFG and RFG close to zero at MAE = 0.44. Similarly, Table \ref{tab:sing_var_fairness_austin} shows the results for RideAustin. Compared to 3D CNN, RF regularizer brings down about 99.5\% RFG (from 62.004 to 0.347) and IF regularizer brings down 80.5\% IFG (from 48.713 to 9.473) while keeping better accuracy. 




\begin{table}[h]
  \caption{FairST compared to baselines for RideAustin demand prediction (single attribute)}
  \label{tab:sing_var_fairness_austin}
\addtolength{\tabcolsep}{-1pt} 
\begin{tabular}{lllllll}
\toprule
             & $\lambda$ & MAE   & \begin{tabular}[c]{@{}l@{}}RFG\\ (race)\end{tabular} & \begin{tabular}[c]{@{}l@{}}IFG\\ (race)\end{tabular} & \begin{tabular}[c]{@{}l@{}}Spearman's\\ rho (race)\end{tabular} \\
\midrule
Ground Truth & /      & /     & 80.120                                               & 59.742                                               & 0.120$^{*}$                                                                                                      \\
HA           & /      & 0.662 & 48.457                                               & 33.550                                               & 0.118$^{*}$                                                                                                       \\
ARIMA        & /      & 0.597 & 82.587                                               & 61.457                                               & 0.117$^{*}$                                                                                                     \\
LSTM\cite{gers1999learning}        & /   & 0.570 & 61.329                                               & 42.101                                               & 0.073                                                                                                      \\
ConvLSTM\cite{xingjian2015convolutional}     & 0.00   & 0.567 & 66.428                                               & 46.534                                               & 0.073                                                                                                     \\
3D CNN       & 0.00   & 0.532 & 62.004                                               & 48.713                                               & 0.051                                                                                                     \\
3D CNN + 1D  & 0.00   & 0.484 & 69.130                                               & 51.048                                               & 0.095                                                                                                    \\
3D CNN + 2D  & 0.00   & 0.482 & 71.309                                               & 50.630                                               & 0.089                                                                                                     \\
FairST       & 0.00   & 0.472 & 76.340                                               & 54.274                                               & 0.073                                                                                                     \\
FairST + RF  & 0.05   & 0.475 & 56.703                                               & 49.092                                               & \textbf{-0.034}                                                                                                   \\
FairST + RF  & 0.80   & 0.524 & \textbf{0.347}                                               & 32.436                                               & -0.059                                                                                                     \\
FairST + RF  & 1.00   & 0.553 & -2.499                                               & 30.327                                               & -0.132$^{*}$                                                                                                      \\
FairST + IF  & 0.06   & \textbf{0.463} & 67.358                                               & 50.357                                               & 0.131$^{*}$                                                                                                     \\
FairST + IF  & 1.20   & 0.515 & -27.397                                              & 9.473                                                & -0.100                                                                                                     \\
FairST + IF  & 2.00   & 0.554 & -38.990                                              & \textbf{0.166}                                              & -0.157$^{**}$                                                    \\
\bottomrule 
\multicolumn{5}{l}{\footnotesize{$^{**}$. Correlation is significant at the 0.05 level. }}\\
\multicolumn{5}{l}{\footnotesize{$^*$. Correlation is significant at the 0.01 level.}}\\
\end{tabular}
\addtolength{\tabcolsep}{1pt} 

\end{table}

\begin{figure*}[h]
  \centering
  \includegraphics[width=0.82\linewidth]{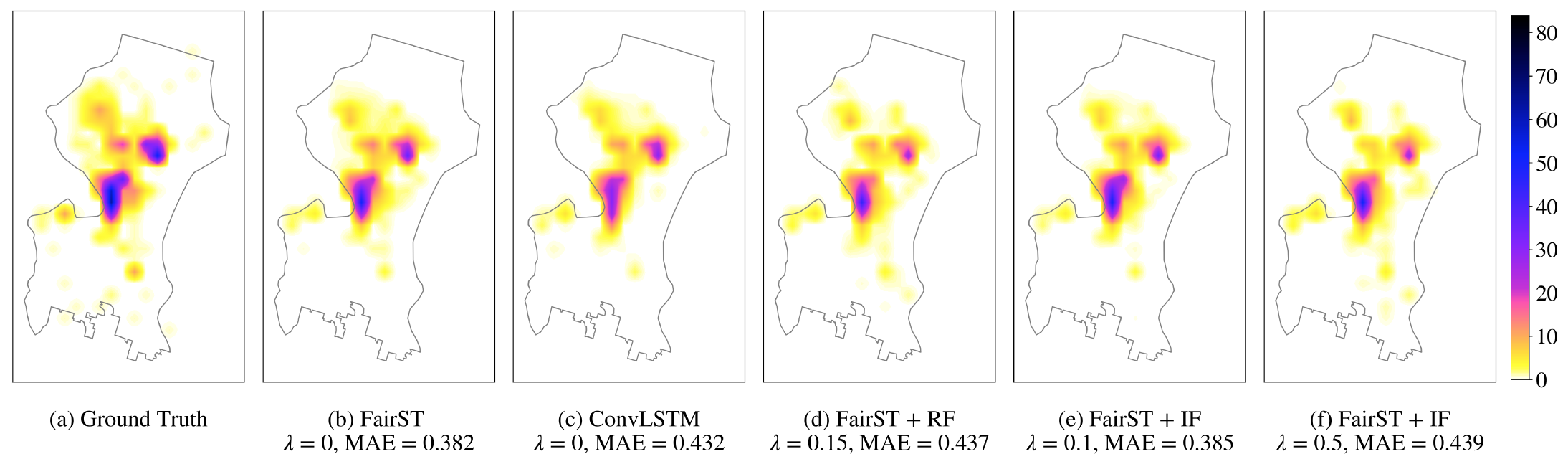}
  \caption{Ground truth vs. predictions heat maps for September 27, 2018 16:00 pm - 17:00 pm. (d), (e), (f) are the predictions from FairST using RF or IF regularizier on multiple sensitive attributes.}\label{fig:heatmap_multivar}
\end{figure*}


In summary, in the single sensitive attribute scenario, FairST is able to achieve an accuracy better than the state-of-the-art baseline models while closing more than 80\% of fairness gap. The proposed fairness regularizers are more effective than baseline fairness regularizers in reducing unfairness.




\subsection{Fair Prediction: Multiple Attributes}

Having demonstrated the effectiveness of closing fairness gaps with IF and RF regularizers on a single sensitive attribute, we now turn to multiple sensitive attributes. We conduct two experiments on Seattle bikeshare dataset using RF loss and IF loss, respectively according to Equation \ref{eq:multivar}. We set $\lambda_{a}$ to be 1.0 for all three attributes, i.e. race, age, and education level. 

\begin{figure}[h]
  \centering
  \includegraphics[width=\linewidth]{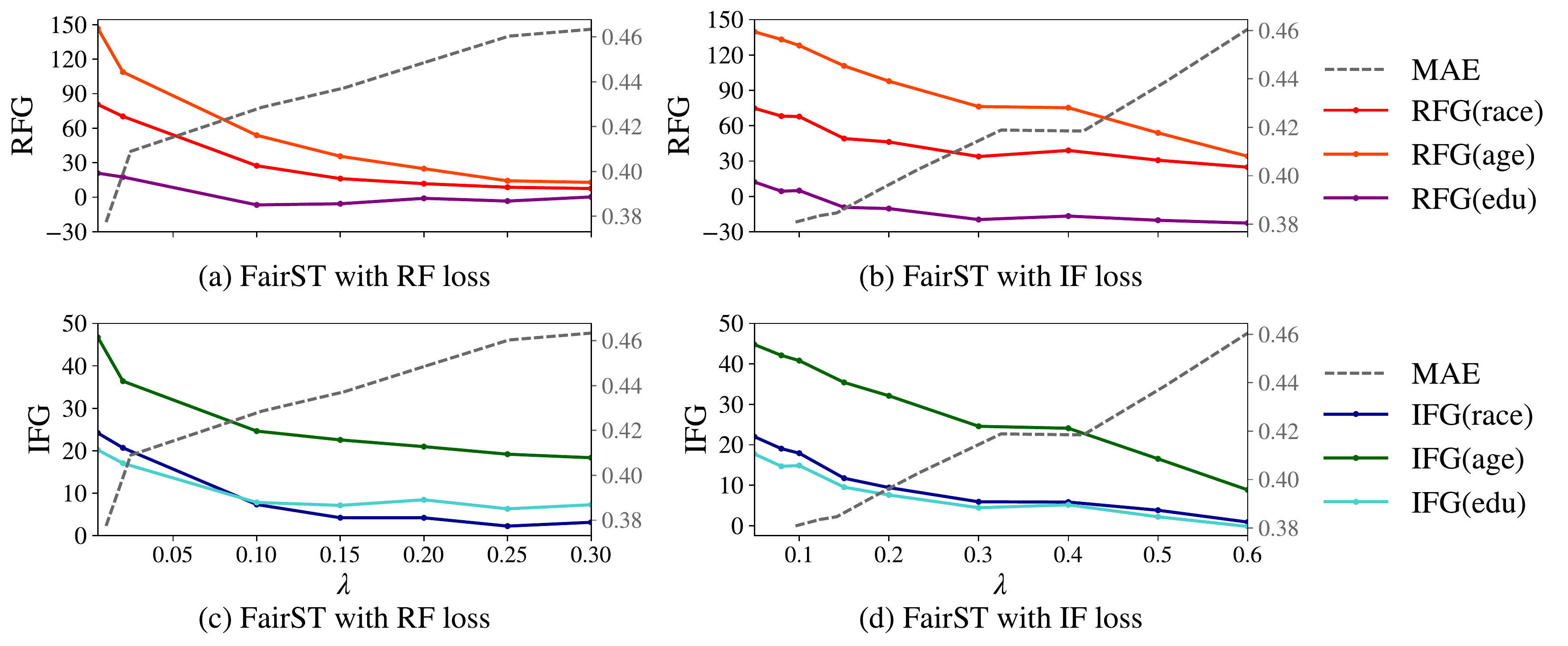}
  \caption{$\lambda$ vs. fairness loss. (a) and (c) show the results of FairST with RF regularizer. (b) and (d) show the results of FairST with IF regularizer.}
  \label{fig:multivar}
\end{figure}

Figure \ref{fig:multivar} shows the results of FairST with RF ((a) and (c)) and IF regularizer ((b) and (d)) evaluated using RFG and IFG. Overall, as $\lambda$ increases, accuracy decreases and fairness increases, indicating that both regularizers consistently help the model to approach equity on multiple sensitive attributes without sacrificing too much accuracy.

We now step back to compare FairST and baselines in terms of accuracy and fairness. Table \ref{tab:mae_fairness} shows the results of FairST with RF regularizer and IF regularizer, denoted by FairST + RF and FairST + IF, with different $\lambda$s. As can be observed, ground truth shows demand gaps between groups, indicating that there were more bikes picked up by whites, young people and college-educated people than the others. There are also significant positive correlations between demand and sensitive attributes (age and education level) as indicated by Spearman's coefficients. Compared to ground truth, all baseline models without fairness consideration amplify inequality in terms of one or more metrics. LSTM achieves good accuracy but drastically enlarges fairness gaps of race and education. ConvLSTM shows better fairness than all baselines in terms of IFG and RFG, but gives higher Spearman's coefficients for race and age than 3D CNN model. FairST with IF or RF regularizer can help reducing inequity in terms of all metrics. For example, compared to ConvLSTM, FairST + RF ($\lambda$ = 0.15) and FairST + IF ($\lambda$ = 0.5) show comparable accuracy but better fairness in terms of all fairness metrics. FairST + IF ($\lambda$ = 0.15) outperforms 3D CNN in both accuracy and fairness.

To understand better how FairST achieves fairness, we visualize the predictions from five different settings as illustrated in Figure \ref{fig:heatmap_multivar}. All five models are capable of learning spatial-temporal dependencies. FairST ($\lambda$ = 0) accurately highlights the hot spots. Compared to ConvLSTM, FairSTs are better at capturing fragmented details around major hot spots. Adding fairness regularizers to FairST preserved the major hot spots but "re-weighted" some values in place and "redistributed" demand from some neighborhoods to others. For example, compared to Figure \ref{fig:heatmap_multivar} (b) which does not consider fairness, Figure \ref{fig:heatmap_multivar} (d) and Figure \ref{fig:heatmap_multivar} (f) tend to capture more demand from the south part of the city where the disadvantaged population concentrates, and less demand from the northwest part of city which is dominated by the wealthy and well-educated population. 



To summarize, in multiple sensitive attributes scenario, FairST is able to reduce unfairness for all three attributes consistently. With selected regularizer weight, FairST outperforms baseline models in both accuracy and fairness. 



\section{Conclusion}

In this paper, we introduced FairST, a fairness-aware spatio-temporal model for predicting new mobility resource demand. Building on 3D convolutional neural network approaches, the model captures spatial-temporal correlations of dynamic new mobility systems. A key feature of FairST is the integration of fairness regularizers to the model to encourage equitable prediction. We also proposed two fairness metrics that measure equity gaps between social groups for urban mobility systems. Experiments on two real-world new mobility datasets demonstrate that FairST is able to close more than 80\% of fairness gap for a single sensitive attribute and at the same time achieve \emph{better} accuracy than state-of-the-art but fairness-oblivious baseline methods. Further experiments show that FairST is able to reduce unfairness for multiple attributes, outperforming baselines in both accuracy and fairness. Future work involves generalizing to arbitrary spatio-temporal prediction problems, considering variant architectures for integrating datasets of heterogeneous dimension, and pre-training features for exogensous variables to save computation time and facilitate data release.

\bibliographystyle{ACM-Reference-Format}

\begin{thebibliography}{47}


\ifx \showCODEN    \undefined \def \showCODEN     #1{\unskip}     \fi
\ifx \showDOI      \undefined \def \showDOI       #1{#1}\fi
\ifx \showISBNx    \undefined \def \showISBNx     #1{\unskip}     \fi
\ifx \showISBNxiii \undefined \def \showISBNxiii  #1{\unskip}     \fi
\ifx \showISSN     \undefined \def \showISSN      #1{\unskip}     \fi
\ifx \showLCCN     \undefined \def \showLCCN      #1{\unskip}     \fi
\ifx \shownote     \undefined \def \shownote      #1{#1}          \fi
\ifx \showarticletitle \undefined \def \showarticletitle #1{#1}   \fi
\ifx \showURL      \undefined \def \showURL       {\relax}        \fi
\providecommand\bibfield[2]{#2}
\providecommand\bibinfo[2]{#2}
\providecommand\natexlab[1]{#1}
\providecommand\showeprint[2][]{arXiv:#2}

\bibitem[\protect\citeauthoryear{Abadi, Barham, Chen, Chen, Davis, Dean, Devin,
  Ghemawat, Irving, Isard, et~al\mbox{.}}{Abadi et~al\mbox{.}}{2016}]%
        {abadi2016tensorflow}
\bibfield{author}{\bibinfo{person}{Mart{\'\i}n Abadi}, \bibinfo{person}{Paul
  Barham}, \bibinfo{person}{Jianmin Chen}, \bibinfo{person}{Zhifeng Chen},
  \bibinfo{person}{Andy Davis}, \bibinfo{person}{Jeffrey Dean},
  \bibinfo{person}{Matthieu Devin}, \bibinfo{person}{Sanjay Ghemawat},
  \bibinfo{person}{Geoffrey Irving}, \bibinfo{person}{Michael Isard},
  {et~al\mbox{.}}} \bibinfo{year}{2016}\natexlab{}.
\newblock \showarticletitle{Tensorflow: a system for large-scale machine
  learning.}. In \bibinfo{booktitle}{\emph{OSDI}}, Vol.~\bibinfo{volume}{16}.
  \bibinfo{pages}{265--283}.
\newblock


\bibitem[\protect\citeauthoryear{Bell and Smyl}{Bell and Smyl}{2018}]%
        {forecastingatUber}
\bibfield{author}{\bibinfo{person}{Franziska Bell} {and}
  \bibinfo{person}{Slawek Smyl}.} \bibinfo{year}{2018}\natexlab{}.
\newblock \showarticletitle{Forecasting at Uber: An Introduction}.
\newblock \bibinfo{journal}{\emph{Uber Engineering}} (\bibinfo{year}{2018}).
\newblock
\urldef\tempurl%
\url{https://eng.uber.com/forecasting-introduction/}
\showURL{%
\tempurl}


\bibitem[\protect\citeauthoryear{Berk, Heidari, Jabbari, Joseph, Kearns,
  Morgenstern, Neel, and Roth}{Berk et~al\mbox{.}}{2017}]%
        {DBLP:journals/corr/BerkHJJKMNR17}
\bibfield{author}{\bibinfo{person}{Richard Berk}, \bibinfo{person}{Hoda
  Heidari}, \bibinfo{person}{Shahin Jabbari}, \bibinfo{person}{Matthew Joseph},
  \bibinfo{person}{Michael~J. Kearns}, \bibinfo{person}{Jamie Morgenstern},
  \bibinfo{person}{Seth Neel}, {and} \bibinfo{person}{Aaron Roth}.}
  \bibinfo{year}{2017}\natexlab{}.
\newblock \showarticletitle{A Convex Framework for Fair Regression}.
\newblock \bibinfo{journal}{\emph{CoRR}}  \bibinfo{volume}{abs/1706.02409}
  (\bibinfo{year}{2017}).
\newblock


\bibitem[\protect\citeauthoryear{Brown}{Brown}{2018}]%
        {brown2018ridehail}
\bibfield{author}{\bibinfo{person}{Anne~Elizabeth Brown}.}
  \bibinfo{year}{2018}\natexlab{}.
\newblock \emph{\bibinfo{title}{Ridehail revolution: Ridehail travel and equity
  in Los Angeles}}.
\newblock \bibinfo{thesistype}{Ph.D. Dissertation}. \bibinfo{school}{UCLA}.
\newblock


\bibitem[\protect\citeauthoryear{Calders, Karim, Kamiran, Ali, and
  Zhang}{Calders et~al\mbox{.}}{2013}]%
        {calders2013controlling}
\bibfield{author}{\bibinfo{person}{Toon Calders}, \bibinfo{person}{Asim Karim},
  \bibinfo{person}{Faisal Kamiran}, \bibinfo{person}{Wasif Ali}, {and}
  \bibinfo{person}{Xiangliang Zhang}.} \bibinfo{year}{2013}\natexlab{}.
\newblock \showarticletitle{Controlling attribute effect in linear regression}.
  In \bibinfo{booktitle}{\emph{Data Mining (ICDM), 2013 IEEE 13th International
  Conference on}}. IEEE, \bibinfo{pages}{71--80}.
\newblock


\bibitem[\protect\citeauthoryear{Carson}{Carson}{2018}]%
        {carson2018lyft}
\bibfield{author}{\bibinfo{person}{Biz Carson}.}
  \bibinfo{year}{2018}\natexlab{}.
\newblock \bibinfo{title}{Lyft doubled rides in 2017 as its rival Uber
  stumbled. Forbes}.
\newblock
\newblock


\bibitem[\protect\citeauthoryear{Delbosc and Currie}{Delbosc and
  Currie}{2011}]%
        {delbosc2011using}
\bibfield{author}{\bibinfo{person}{Alexa Delbosc} {and} \bibinfo{person}{Graham
  Currie}.} \bibinfo{year}{2011}\natexlab{}.
\newblock \showarticletitle{Using Lorenz curves to assess public transport
  equity}.
\newblock \bibinfo{journal}{\emph{Journal of Transport Geography}}
  \bibinfo{volume}{19}, \bibinfo{number}{6} (\bibinfo{year}{2011}),
  \bibinfo{pages}{1252--1259}.
\newblock


\bibitem[\protect\citeauthoryear{Dwork, Hardt, Pitassi, Reingold, and
  Zemel}{Dwork et~al\mbox{.}}{2012}]%
        {Dwork:2012:FTA:2090236.2090255}
\bibfield{author}{\bibinfo{person}{Cynthia Dwork}, \bibinfo{person}{Moritz
  Hardt}, \bibinfo{person}{Toniann Pitassi}, \bibinfo{person}{Omer Reingold},
  {and} \bibinfo{person}{Richard Zemel}.} \bibinfo{year}{2012}\natexlab{}.
\newblock \showarticletitle{Fairness Through Awareness}. In
  \bibinfo{booktitle}{\emph{Proceedings of the 3rd Innovations in Theoretical
  Computer Science Conference}} \emph{(\bibinfo{series}{ITCS '12})}.
  \bibinfo{pages}{214--226}.
\newblock
\showISBNx{978-1-4503-1115-1}


\bibitem[\protect\citeauthoryear{El-Assi, Mahmoud, and Habib}{El-Assi
  et~al\mbox{.}}{2017}]%
        {el2017effects}
\bibfield{author}{\bibinfo{person}{Wafic El-Assi},
  \bibinfo{person}{Mohamed~Salah Mahmoud}, {and}
  \bibinfo{person}{Khandker~Nurul Habib}.} \bibinfo{year}{2017}\natexlab{}.
\newblock \showarticletitle{Effects of built environment and weather on bike
  sharing demand: a station level analysis of commercial bike sharing in
  Toronto}.
\newblock \bibinfo{journal}{\emph{Transportation}} \bibinfo{volume}{44},
  \bibinfo{number}{3} (\bibinfo{year}{2017}), \bibinfo{pages}{589--613}.
\newblock


\bibitem[\protect\citeauthoryear{Feldman, Friedler, Moeller, Scheidegger, and
  Venkatasubramanian}{Feldman et~al\mbox{.}}{2015}]%
        {Feldman:2015:CRD:2783258.2783311}
\bibfield{author}{\bibinfo{person}{Michael Feldman},
  \bibinfo{person}{Sorelle~A. Friedler}, \bibinfo{person}{John Moeller},
  \bibinfo{person}{Carlos Scheidegger}, {and} \bibinfo{person}{Suresh
  Venkatasubramanian}.} \bibinfo{year}{2015}\natexlab{}.
\newblock \showarticletitle{Certifying and Removing Disparate Impact}. In
  \bibinfo{booktitle}{\emph{Proceedings of the 21th ACM SIGKDD International
  Conference on Knowledge Discovery and Data Mining}}
  \emph{(\bibinfo{series}{KDD '15})}. \bibinfo{pages}{259--268}.
\newblock
\showISBNx{978-1-4503-3664-2}


\bibitem[\protect\citeauthoryear{Frade and Ribeiro}{Frade and Ribeiro}{2014}]%
        {frade2014bicycle}
\bibfield{author}{\bibinfo{person}{In{\^e}s Frade} {and}
  \bibinfo{person}{Anabela Ribeiro}.} \bibinfo{year}{2014}\natexlab{}.
\newblock \showarticletitle{Bicycle sharing systems demand}.
\newblock \bibinfo{journal}{\emph{Procedia-Social and Behavioral Sciences}}
  \bibinfo{volume}{111} (\bibinfo{year}{2014}), \bibinfo{pages}{518--527}.
\newblock


\bibitem[\protect\citeauthoryear{Ge, Knittel, MacKenzie, and Zoepf}{Ge
  et~al\mbox{.}}{2016}]%
        {ge2016racial}
\bibfield{author}{\bibinfo{person}{Yanbo Ge}, \bibinfo{person}{Christopher~R
  Knittel}, \bibinfo{person}{Don MacKenzie}, {and} \bibinfo{person}{Stephen
  Zoepf}.} \bibinfo{year}{2016}\natexlab{}.
\newblock \bibinfo{booktitle}{\emph{Racial and gender discrimination in
  transportation network companies}}.
\newblock \bibinfo{type}{{T}echnical {R}eport}. \bibinfo{institution}{National
  Bureau of Economic Research}.
\newblock


\bibitem[\protect\citeauthoryear{Gers, Schmidhuber, and Cummins}{Gers
  et~al\mbox{.}}{1999}]%
        {gers1999learning}
\bibfield{author}{\bibinfo{person}{Felix~A Gers}, \bibinfo{person}{J{\"u}rgen
  Schmidhuber}, {and} \bibinfo{person}{Fred Cummins}.}
  \bibinfo{year}{1999}\natexlab{}.
\newblock \showarticletitle{Learning to forget: Continual prediction with
  LSTM}.
\newblock  (\bibinfo{year}{1999}).
\newblock


\bibitem[\protect\citeauthoryear{Glymour and Herington}{Glymour and
  Herington}{2019}]%
        {Glymour:2019:MBM:3287560.3287573}
\bibfield{author}{\bibinfo{person}{Bruce Glymour} {and}
  \bibinfo{person}{Jonathan Herington}.} \bibinfo{year}{2019}\natexlab{}.
\newblock \showarticletitle{Measuring the Biases That Matter: The Ethical and
  Casual Foundations for Measures of Fairness in Algorithms}. In
  \bibinfo{booktitle}{\emph{Proceedings of the Conference on Fairness,
  Accountability, and Transparency}} \emph{(\bibinfo{series}{FAT* '19})}.
  \bibinfo{pages}{269--278}.
\newblock
\showISBNx{978-1-4503-6125-5}


\bibitem[\protect\citeauthoryear{Hardt, Price, and Srebro}{Hardt
  et~al\mbox{.}}{2016}]%
        {Hardt:2016:EOS:3157382.3157469}
\bibfield{author}{\bibinfo{person}{Moritz Hardt}, \bibinfo{person}{Eric Price},
  {and} \bibinfo{person}{Nathan Srebro}.} \bibinfo{year}{2016}\natexlab{}.
\newblock \showarticletitle{Equality of Opportunity in Supervised Learning}. In
  \bibinfo{booktitle}{\emph{Proceedings of the 30th International Conference on
  Neural Information Processing Systems}} \emph{(\bibinfo{series}{NIPS'16})}.
  \bibinfo{pages}{3323--3331}.
\newblock
\showISBNx{978-1-5108-3881-9}


\bibitem[\protect\citeauthoryear{Hauke and Kossowski}{Hauke and
  Kossowski}{2011}]%
        {hauke2011comparison}
\bibfield{author}{\bibinfo{person}{Jan Hauke} {and} \bibinfo{person}{Tomasz
  Kossowski}.} \bibinfo{year}{2011}\natexlab{}.
\newblock \showarticletitle{Comparison of values of Pearson's and Spearman's
  correlation coefficients on the same sets of data}.
\newblock \bibinfo{journal}{\emph{Quaestiones geographicae}}
  \bibinfo{volume}{30}, \bibinfo{number}{2} (\bibinfo{year}{2011}),
  \bibinfo{pages}{87--93}.
\newblock


\bibitem[\protect\citeauthoryear{Hosford and Winters}{Hosford and
  Winters}{2018}]%
        {hosford2018public}
\bibfield{author}{\bibinfo{person}{Kate Hosford} {and} \bibinfo{person}{Meghan
  Winters}.} \bibinfo{year}{2018}\natexlab{}.
\newblock \showarticletitle{Who are public bicycle share programs serving? An
  evaluation of the equity of spatial access to bicycle share service areas in
  Canadian cities}.
\newblock \bibinfo{journal}{\emph{Transportation research record}}
  (\bibinfo{year}{2018}), \bibinfo{pages}{0361198118783107}.
\newblock


\bibitem[\protect\citeauthoryear{Hughes and MacKenzie}{Hughes and
  MacKenzie}{2016}]%
        {hughes2016transportation}
\bibfield{author}{\bibinfo{person}{Ryan Hughes} {and} \bibinfo{person}{Don
  MacKenzie}.} \bibinfo{year}{2016}\natexlab{}.
\newblock \showarticletitle{Transportation network company wait times in
  Greater Seattle, and relationship to socioeconomic indicators}.
\newblock \bibinfo{journal}{\emph{Journal of Transport Geography}}
  \bibinfo{volume}{56} (\bibinfo{year}{2016}), \bibinfo{pages}{36--44}.
\newblock


\bibitem[\protect\citeauthoryear{Hutchinson and Mitchell}{Hutchinson and
  Mitchell}{2019}]%
        {Hutchinson:2019:YTF:3287560.3287600}
\bibfield{author}{\bibinfo{person}{Ben Hutchinson} {and}
  \bibinfo{person}{Margaret Mitchell}.} \bibinfo{year}{2019}\natexlab{}.
\newblock \showarticletitle{50 Years of Test (Un)Fairness: Lessons for Machine
  Learning}. In \bibinfo{booktitle}{\emph{Proceedings of the Conference on
  Fairness, Accountability, and Transparency}} \emph{(\bibinfo{series}{FAT*
  '19})}. \bibinfo{pages}{49--58}.
\newblock
\showISBNx{978-1-4503-6125-5}


\bibitem[\protect\citeauthoryear{Ji, Xu, Yang, and Yu}{Ji
  et~al\mbox{.}}{2013}]%
        {ji20133d}
\bibfield{author}{\bibinfo{person}{Shuiwang Ji}, \bibinfo{person}{Wei Xu},
  \bibinfo{person}{Ming Yang}, {and} \bibinfo{person}{Kai Yu}.}
  \bibinfo{year}{2013}\natexlab{}.
\newblock \showarticletitle{3D convolutional neural networks for human action
  recognition}.
\newblock \bibinfo{journal}{\emph{IEEE transactions on pattern analysis and
  machine intelligence}} \bibinfo{volume}{35}, \bibinfo{number}{1}
  (\bibinfo{year}{2013}), \bibinfo{pages}{221--231}.
\newblock


\bibitem[\protect\citeauthoryear{Kamiran and Calders}{Kamiran and
  Calders}{2009}]%
        {kamiran2009classifying}
\bibfield{author}{\bibinfo{person}{Faisal Kamiran} {and} \bibinfo{person}{Toon
  Calders}.} \bibinfo{year}{2009}\natexlab{}.
\newblock \showarticletitle{Classifying without discriminating}. In
  \bibinfo{booktitle}{\emph{Computer, Control and Communication, 2009. IC4
  2009. 2nd International Conference on}}. IEEE, \bibinfo{pages}{1--6}.
\newblock


\bibitem[\protect\citeauthoryear{Komiyama, Takeda, Honda, and Shimao}{Komiyama
  et~al\mbox{.}}{2018}]%
        {pmlr-v80-komiyama18a}
\bibfield{author}{\bibinfo{person}{Junpei Komiyama}, \bibinfo{person}{Akiko
  Takeda}, \bibinfo{person}{Junya Honda}, {and} \bibinfo{person}{Hajime
  Shimao}.} \bibinfo{year}{2018}\natexlab{}.
\newblock \showarticletitle{Nonconvex Optimization for Regression with Fairness
  Constraints}. In \bibinfo{booktitle}{\emph{Proceedings of the 35th
  International Conference on Machine Learning}}
  \emph{(\bibinfo{series}{Proceedings of Machine Learning Research})},
  \bibfield{editor}{\bibinfo{person}{Jennifer Dy} {and}
  \bibinfo{person}{Andreas Krause}} (Eds.), Vol.~\bibinfo{volume}{80}.
  \bibinfo{publisher}{PMLR}, \bibinfo{address}{Stockholmsmässan, Stockholm
  Sweden}, \bibinfo{pages}{2737--2746}.
\newblock


\bibitem[\protect\citeauthoryear{Li, Zhang, Sun, and Liu}{Li
  et~al\mbox{.}}{2018a}]%
        {li2018free}
\bibfield{author}{\bibinfo{person}{Xuefeng Li}, \bibinfo{person}{Yong Zhang},
  \bibinfo{person}{Li Sun}, {and} \bibinfo{person}{Qiyang Liu}.}
  \bibinfo{year}{2018}\natexlab{a}.
\newblock \showarticletitle{Free-Floating Bike Sharing in Jiangsu: Users'
  Behaviors and Influencing Factors}.
\newblock \bibinfo{journal}{\emph{Energies}} \bibinfo{volume}{11},
  \bibinfo{number}{7} (\bibinfo{year}{2018}), \bibinfo{pages}{1664}.
\newblock


\bibitem[\protect\citeauthoryear{Li, Zheng, and Yang}{Li
  et~al\mbox{.}}{2018b}]%
        {Li:2018:DBR:3219819.3220110}
\bibfield{author}{\bibinfo{person}{Yexin Li}, \bibinfo{person}{Yu Zheng}, {and}
  \bibinfo{person}{Qiang Yang}.} \bibinfo{year}{2018}\natexlab{b}.
\newblock \showarticletitle{Dynamic Bike Reposition: A Spatio-Temporal
  Reinforcement Learning Approach}. In \bibinfo{booktitle}{\emph{Proceedings of
  the 24th ACM SIGKDD International Conference on Knowledge Discovery and Data
  Mining}} \emph{(\bibinfo{series}{KDD '18})}. \bibinfo{pages}{1724--1733}.
\newblock
\showISBNx{978-1-4503-5552-0}


\bibitem[\protect\citeauthoryear{Li, Zheng, Zhang, and Chen}{Li
  et~al\mbox{.}}{2015}]%
        {Li:2015:TPB:2820783.2820837}
\bibfield{author}{\bibinfo{person}{Yexin Li}, \bibinfo{person}{Yu Zheng},
  \bibinfo{person}{Huichu Zhang}, {and} \bibinfo{person}{Lei Chen}.}
  \bibinfo{year}{2015}\natexlab{}.
\newblock \showarticletitle{Traffic Prediction in a Bike-sharing System}. In
  \bibinfo{booktitle}{\emph{Proceedings of the 23rd SIGSPATIAL International
  Conference on Advances in Geographic Information Systems}}
  \emph{(\bibinfo{series}{SIGSPATIAL '15})}. Article \bibinfo{articleno}{33},
  \bibinfo{numpages}{10}~pages.
\newblock
\showISBNx{978-1-4503-3967-4}


\bibitem[\protect\citeauthoryear{Liu, Wu, Wang, and Tan}{Liu
  et~al\mbox{.}}{2016}]%
        {liu2016predicting}
\bibfield{author}{\bibinfo{person}{Qiang Liu}, \bibinfo{person}{Shu Wu},
  \bibinfo{person}{Liang Wang}, {and} \bibinfo{person}{Tieniu Tan}.}
  \bibinfo{year}{2016}\natexlab{}.
\newblock \showarticletitle{Predicting the Next Location: A Recurrent Model
  with Spatial and Temporal Contexts.}. In \bibinfo{booktitle}{\emph{AAAI}}.
  \bibinfo{pages}{194--200}.
\newblock


\bibitem[\protect\citeauthoryear{Maas, Hannun, and Ng}{Maas
  et~al\mbox{.}}{2013}]%
        {maas2013rectifier}
\bibfield{author}{\bibinfo{person}{Andrew~L Maas}, \bibinfo{person}{Awni~Y
  Hannun}, {and} \bibinfo{person}{Andrew~Y Ng}.}
  \bibinfo{year}{2013}\natexlab{}.
\newblock \showarticletitle{Rectifier nonlinearities improve neural network
  acoustic models}. In \bibinfo{booktitle}{\emph{Proc. icml}},
  Vol.~\bibinfo{volume}{30}. \bibinfo{pages}{3}.
\newblock


\bibitem[\protect\citeauthoryear{McNeil, Dill, MacArthur, Broach, and
  Howland}{McNeil et~al\mbox{.}}{2017}]%
        {mcneil2017breaking}
\bibfield{author}{\bibinfo{person}{Nathan McNeil}, \bibinfo{person}{Jennifer
  Dill}, \bibinfo{person}{John MacArthur}, \bibinfo{person}{Joseph Broach},
  {and} \bibinfo{person}{Steven Howland}.} \bibinfo{year}{2017}\natexlab{}.
\newblock \showarticletitle{Breaking Barriers to Bike Share: Insights from
  Residents of Traditionally Underserved Neighborhoods}.
\newblock  (\bibinfo{year}{2017}).
\newblock


\bibitem[\protect\citeauthoryear{Mooney, Hosford, Howe, Yan, Winters, Bassok,
  and Hirsch}{Mooney et~al\mbox{.}}{2019}]%
        {mooney2019freedom}
\bibfield{author}{\bibinfo{person}{Stephen~J Mooney}, \bibinfo{person}{Kate
  Hosford}, \bibinfo{person}{Bill Howe}, \bibinfo{person}{An Yan},
  \bibinfo{person}{Meghan Winters}, \bibinfo{person}{Alon Bassok}, {and}
  \bibinfo{person}{Jana~A Hirsch}.} \bibinfo{year}{2019}\natexlab{}.
\newblock \showarticletitle{Freedom from the station: Spatial equity in access
  to dockless bike share}.
\newblock \bibinfo{journal}{\emph{Journal of Transport Geography}}
  \bibinfo{volume}{74} (\bibinfo{year}{2019}), \bibinfo{pages}{91--96}.
\newblock


\bibitem[\protect\citeauthoryear{Shen, Liang, Ouyang, Liu, Zheng, and
  Carley}{Shen et~al\mbox{.}}{2018}]%
        {shen2018stepdeep}
\bibfield{author}{\bibinfo{person}{Bilong Shen}, \bibinfo{person}{Xiaodan
  Liang}, \bibinfo{person}{Yufeng Ouyang}, \bibinfo{person}{Miaofeng Liu},
  \bibinfo{person}{Weimin Zheng}, {and} \bibinfo{person}{Kathleen~M Carley}.}
  \bibinfo{year}{2018}\natexlab{}.
\newblock \showarticletitle{StepDeep: A Novel Spatial-temporal Mobility Event
  Prediction Framework based on Deep Neural Network}. In
  \bibinfo{booktitle}{\emph{Proceedings of the 24th ACM SIGKDD International
  Conference on Knowledge Discovery \& Data Mining}}. ACM,
  \bibinfo{pages}{724--733}.
\newblock


\bibitem[\protect\citeauthoryear{SimplyAnalytics}{SimplyAnalytics}{2018}]%
        {EASI2018}
\bibfield{author}{\bibinfo{person}{SimplyAnalytics}.}
  \bibinfo{year}{2018}\natexlab{}.
\newblock \bibinfo{title}{EASI/MRI Census US}.
\newblock
\newblock
\urldef\tempurl%
\url{SimplyAnalytics database}
\showURL{%
Retrieved November 2, 2018 from \tempurl}


\bibitem[\protect\citeauthoryear{Stark and Diakopoulos}{Stark and
  Diakopoulos}{2016}]%
        {stark2016uber}
\bibfield{author}{\bibinfo{person}{Jennifer Stark} {and}
  \bibinfo{person}{Nicholas Diakopoulos}.} \bibinfo{year}{2016}\natexlab{}.
\newblock \showarticletitle{Uber seems to offer better service in areas with
  more white people. That raises some tough questions}.
\newblock \bibinfo{journal}{\emph{The Washington Post}} (\bibinfo{year}{2016}).
\newblock


\bibitem[\protect\citeauthoryear{Szegedy, Liu, Jia, Sermanet, Reed, Anguelov,
  Erhan, Vanhoucke, and Rabinovich}{Szegedy et~al\mbox{.}}{2015}]%
        {szegedy2015going}
\bibfield{author}{\bibinfo{person}{Christian Szegedy}, \bibinfo{person}{Wei
  Liu}, \bibinfo{person}{Yangqing Jia}, \bibinfo{person}{Pierre Sermanet},
  \bibinfo{person}{Scott Reed}, \bibinfo{person}{Dragomir Anguelov},
  \bibinfo{person}{Dumitru Erhan}, \bibinfo{person}{Vincent Vanhoucke}, {and}
  \bibinfo{person}{Andrew Rabinovich}.} \bibinfo{year}{2015}\natexlab{}.
\newblock \showarticletitle{Going deeper with convolutions}. In
  \bibinfo{booktitle}{\emph{Proceedings of the IEEE conference on computer
  vision and pattern recognition}}. \bibinfo{pages}{1--9}.
\newblock


\bibitem[\protect\citeauthoryear{Tran, Bourdev, Fergus, Torresani, and
  Paluri}{Tran et~al\mbox{.}}{2015}]%
        {Tran:2015:LSF:2919332.2919929}
\bibfield{author}{\bibinfo{person}{Du Tran}, \bibinfo{person}{Lubomir Bourdev},
  \bibinfo{person}{Rob Fergus}, \bibinfo{person}{Lorenzo Torresani}, {and}
  \bibinfo{person}{Manohar Paluri}.} \bibinfo{year}{2015}\natexlab{}.
\newblock \showarticletitle{Learning Spatiotemporal Features with 3D
  Convolutional Networks}. In \bibinfo{booktitle}{\emph{Proceedings of the 2015
  IEEE International Conference on Computer Vision (ICCV)}}
  \emph{(\bibinfo{series}{ICCV '15})}. \bibinfo{pages}{4489--4497}.
\newblock
\showISBNx{978-1-4673-8391-2}


\bibitem[\protect\citeauthoryear{Ursaki and Aultman-Hall}{Ursaki and
  Aultman-Hall}{2016}]%
        {ursaki2016quantifying}
\bibfield{author}{\bibinfo{person}{Julia Ursaki} {and} \bibinfo{person}{Lisa
  Aultman-Hall}.} \bibinfo{year}{2016}\natexlab{}.
\newblock \showarticletitle{Quantifying the equity of bikeshare access in US
  cities}. In \bibinfo{booktitle}{\emph{95th Annual Meeting of the
  Transportation Research Board, Washington, DC}}.
\newblock


\bibitem[\protect\citeauthoryear{Vogel, Greiser, and Mattfeld}{Vogel
  et~al\mbox{.}}{2011}]%
        {vogel2011understanding}
\bibfield{author}{\bibinfo{person}{Patrick Vogel}, \bibinfo{person}{Torsten
  Greiser}, {and} \bibinfo{person}{Dirk~Christian Mattfeld}.}
  \bibinfo{year}{2011}\natexlab{}.
\newblock \showarticletitle{Understanding bike-sharing systems using data
  mining: Exploring activity patterns}.
\newblock \bibinfo{journal}{\emph{Procedia-Social and Behavioral Sciences}}
  \bibinfo{volume}{20} (\bibinfo{year}{2011}), \bibinfo{pages}{514--523}.
\newblock


\bibitem[\protect\citeauthoryear{Wang, Cao, Li, and Ye}{Wang
  et~al\mbox{.}}{2017}]%
        {wang2017deepsd}
\bibfield{author}{\bibinfo{person}{Dong Wang}, \bibinfo{person}{Wei Cao},
  \bibinfo{person}{Jian Li}, {and} \bibinfo{person}{Jieping Ye}.}
  \bibinfo{year}{2017}\natexlab{}.
\newblock \showarticletitle{DeepSD: supply-demand prediction for online
  car-hailing services using deep neural networks}. In
  \bibinfo{booktitle}{\emph{2017 IEEE 33rd International Conference on Data
  Engineering (ICDE)}}. IEEE, \bibinfo{pages}{243--254}.
\newblock


\bibitem[\protect\citeauthoryear{Wang and Mu}{Wang and Mu}{2018}]%
        {wang2018spatial}
\bibfield{author}{\bibinfo{person}{Mingshu Wang} {and} \bibinfo{person}{Lan
  Mu}.} \bibinfo{year}{2018}\natexlab{}.
\newblock \showarticletitle{Spatial disparities of Uber accessibility: An
  exploratory analysis in Atlanta, USA}.
\newblock \bibinfo{journal}{\emph{Computers, Environment and Urban Systems}}
  \bibinfo{volume}{67} (\bibinfo{year}{2018}), \bibinfo{pages}{169--175}.
\newblock


\bibitem[\protect\citeauthoryear{Xingjian, Chen, Wang, Yeung, Wong, and
  Woo}{Xingjian et~al\mbox{.}}{2015}]%
        {xingjian2015convolutional}
\bibfield{author}{\bibinfo{person}{Shi Xingjian}, \bibinfo{person}{Zhourong
  Chen}, \bibinfo{person}{Hao Wang}, \bibinfo{person}{Dit-Yan Yeung},
  \bibinfo{person}{Wai-Kin Wong}, {and} \bibinfo{person}{Wang-chun Woo}.}
  \bibinfo{year}{2015}\natexlab{}.
\newblock \showarticletitle{Convolutional LSTM network: A machine learning
  approach for precipitation nowcasting}. In \bibinfo{booktitle}{\emph{Advances
  in neural information processing systems}}. \bibinfo{pages}{802--810}.
\newblock


\bibitem[\protect\citeauthoryear{Xu, Ji, and Liu}{Xu et~al\mbox{.}}{2018}]%
        {xu2018station}
\bibfield{author}{\bibinfo{person}{Chengcheng Xu}, \bibinfo{person}{Junyi Ji},
  {and} \bibinfo{person}{Pan Liu}.} \bibinfo{year}{2018}\natexlab{}.
\newblock \showarticletitle{The station-free sharing bike demand forecasting
  with a deep learning approach and large-scale datasets}.
\newblock \bibinfo{journal}{\emph{Transportation Research Part C: Emerging
  Technologies}}  \bibinfo{volume}{95} (\bibinfo{year}{2018}),
  \bibinfo{pages}{47--60}.
\newblock


\bibitem[\protect\citeauthoryear{Yao, Wu, Ke, Tang, Jia, Lu, Gong, Ye, and
  Li}{Yao et~al\mbox{.}}{2018}]%
        {Yao2018DeepMS}
\bibfield{author}{\bibinfo{person}{Huaxiu Yao}, \bibinfo{person}{Fei Wu},
  \bibinfo{person}{Jintao Ke}, \bibinfo{person}{Xianfeng Tang},
  \bibinfo{person}{Yitian Jia}, \bibinfo{person}{Siyu Lu},
  \bibinfo{person}{Pinghua Gong}, \bibinfo{person}{Jieping Ye}, {and}
  \bibinfo{person}{Zhenhui Li}.} \bibinfo{year}{2018}\natexlab{}.
\newblock \showarticletitle{Deep Multi-View Spatial-Temporal Network for Taxi
  Demand Prediction}. In \bibinfo{booktitle}{\emph{AAAI}}.
\newblock


\bibitem[\protect\citeauthoryear{Yoon, Pinelli, and Calabrese}{Yoon
  et~al\mbox{.}}{2012}]%
        {yoon2012cityride}
\bibfield{author}{\bibinfo{person}{Ji~Won Yoon}, \bibinfo{person}{Fabio
  Pinelli}, {and} \bibinfo{person}{Francesco Calabrese}.}
  \bibinfo{year}{2012}\natexlab{}.
\newblock \showarticletitle{Cityride: a predictive bike sharing journey
  advisor}. In \bibinfo{booktitle}{\emph{Mobile Data Management (MDM), 2012
  IEEE 13th International Conference on}}. IEEE, \bibinfo{pages}{306--311}.
\newblock


\bibitem[\protect\citeauthoryear{Yuan, Zhou, and Yang}{Yuan
  et~al\mbox{.}}{2018}]%
        {yuan2018hetero}
\bibfield{author}{\bibinfo{person}{Zhuoning Yuan}, \bibinfo{person}{Xun Zhou},
  {and} \bibinfo{person}{Tianbao Yang}.} \bibinfo{year}{2018}\natexlab{}.
\newblock \showarticletitle{Hetero-ConvLSTM: A Deep Learning Approach to
  Traffic Accident Prediction on Heterogeneous Spatio-Temporal Data}. In
  \bibinfo{booktitle}{\emph{Proceedings of the 24th ACM SIGKDD International
  Conference on Knowledge Discovery \& Data Mining}}. ACM,
  \bibinfo{pages}{984--992}.
\newblock


\bibitem[\protect\citeauthoryear{Zafar, Valera, Rodriguez, and Gummadi}{Zafar
  et~al\mbox{.}}{2015}]%
        {zafar2015fairness}
\bibfield{author}{\bibinfo{person}{Muhammad~Bilal Zafar},
  \bibinfo{person}{Isabel Valera}, \bibinfo{person}{Manuel~Gomez Rodriguez},
  {and} \bibinfo{person}{Krishna~P Gummadi}.} \bibinfo{year}{2015}\natexlab{}.
\newblock \showarticletitle{Fairness constraints: Mechanisms for fair
  classification}.
\newblock \bibinfo{journal}{\emph{arXiv preprint arXiv:1507.05259}}
  (\bibinfo{year}{2015}).
\newblock


\bibitem[\protect\citeauthoryear{Zemel, Wu, Swersky, Pitassi, and Dwork}{Zemel
  et~al\mbox{.}}{2013}]%
        {Zemel:2013:LFR:3042817.3042973}
\bibfield{author}{\bibinfo{person}{Richard Zemel}, \bibinfo{person}{Yu Wu},
  \bibinfo{person}{Kevin Swersky}, \bibinfo{person}{Toniann Pitassi}, {and}
  \bibinfo{person}{Cynthia Dwork}.} \bibinfo{year}{2013}\natexlab{}.
\newblock \showarticletitle{Learning Fair Representations}. In
  \bibinfo{booktitle}{\emph{Proceedings of the 30th International Conference on
  International Conference on Machine Learning - Volume 28}}
  \emph{(\bibinfo{series}{ICML'13})}. \bibinfo{pages}{III--325--III--333}.
\newblock


\bibitem[\protect\citeauthoryear{Zhang, Zheng, Qi, Li, Yi, and Li}{Zhang
  et~al\mbox{.}}{2018}]%
        {ZHANG2018147}
\bibfield{author}{\bibinfo{person}{Junbo Zhang}, \bibinfo{person}{Yu Zheng},
  \bibinfo{person}{Dekang Qi}, \bibinfo{person}{Ruiyuan Li},
  \bibinfo{person}{Xiuwen Yi}, {and} \bibinfo{person}{Tianrui Li}.}
  \bibinfo{year}{2018}\natexlab{}.
\newblock \showarticletitle{Predicting citywide crowd flows using deep
  spatio-temporal residual networks}.
\newblock \bibinfo{journal}{\emph{Artificial Intelligence}}
  \bibinfo{volume}{259} (\bibinfo{year}{2018}), \bibinfo{pages}{147 -- 166}.
\newblock
\showISSN{0004-3702}
\urldef\tempurl%
\url{https://doi.org/10.1016/j.artint.2018.03.002}
\showDOI{\tempurl}


\bibitem[\protect\citeauthoryear{Zheng, Capra, Wolfson, and Yang}{Zheng
  et~al\mbox{.}}{2014}]%
        {zheng2014urban}
\bibfield{author}{\bibinfo{person}{Yu Zheng}, \bibinfo{person}{Licia Capra},
  \bibinfo{person}{Ouri Wolfson}, {and} \bibinfo{person}{Hai Yang}.}
  \bibinfo{year}{2014}\natexlab{}.
\newblock \showarticletitle{Urban computing: concepts, methodologies, and
  applications}.
\newblock \bibinfo{journal}{\emph{ACM Transactions on Intelligent Systems and
  Technology (TIST)}} \bibinfo{volume}{5}, \bibinfo{number}{3}
  (\bibinfo{year}{2014}), \bibinfo{pages}{38}.
\newblock


\end{thebibliography}


\newpage
%
\appendix

\end{document}